\documentstyle[11pt,aaspp]{article}
\begin{document}
\newbox\grsign \setbox\grsign=\hbox{$>$} \newdimen\grdimen \grdimen=\ht\grsign
\newbox\simlessbox \newbox\simgreatbox
\setbox\simgreatbox=\hbox{\raise.5ex\hbox{$>$}\llap
     {\lower.5ex\hbox{$\sim$}}}\ht1=\grdimen\dp1=0pt
\setbox\simlessbox=\hbox{\raise.5ex\hbox{$<$}\llap
     {\lower.5ex\hbox{$\sim$}}}\ht2=\grdimen\dp2=0pt
\def\gtorder{\mathrel{\copy\simgreatbox}}
\def\ltorder{\mathrel{\copy\simlessbox}}
\def\simgreat{\mathrel{\copy\simgreatbox}}
\def\simless{\mathrel{\copy\simlessbox}}

%input macro
%at the start of a tex file. Note that different printers may or
%may not automatically include an offset---if the text is off to
%one side adjust the commands \hoffset and \voffset by adding or
%removing a comment sign %.
\def\chaphead{}

\def\hut{Hubble type\ }
\def\vc{V$_{\rm C}$\ }
\def\mb{M$_{\rm B}$\ }
\def\av{A$_{\rm V}$\ }
\def\lamlam{$\lambda\lambda$}

\def\deg{$^\circ$}
\def\degrees{$^\circ$}
\def\Vlasov{collisionless Boltzmann\ }
\def\lsls{\ll}
\def\grgr{\gg}
\def\erf{\mathop{\rm erf}\nolimits} %error function
\def\eqv{\equiv}
\def\real{\Re e}
\def\imag{\Im m}
\def\ctrline#1{\centerline{#1}}
\def\spose#1{\hbox to 0pt{#1\hss}}
     
\def\={\overline}
\def\sections{\S}
\newcount\notenumber
\notenumber=1
\newcount\eqnumber
\eqnumber=1
\newcount\fignumber
\fignumber=1
\newbox\abstr
\newbox\figca     
\def\yyskip{\penalty-100\vskip6pt plus6pt minus4pt}
     
%\numberpara produces numbered paragraphs with extra space and no indentation
\def\numberpara{\yyskip\noindent}
     
\def\km{{\rm\,km}}
\def\kms{{\rm\ km\ s$^{-1}$}}
\def\kpc{{\rm\,kpc}}
\def\mpc{{\rm\,Mpc}}
\def\etal{{\it et al. }}
\def\eg{{\it e.g. }}
\def\ie{{\it i.e. }}
\def\cf{{\it cf. }}
\def\msun{{\rm\,M_\odot}}
\def\lsun{{\rm\,L_\odot}}
\def\rsun{{\rm\,R_\odot}}
\def\pc{{\rm\,pc}}
\def\cm{{\rm\,cm}}
\def\yr{{\rm\,yr}}
\def\au{{\rm\,AU}}
\def\AU{{\rm\,AU}}
\def\gm{{\rm\,g}}
\def\s{{\rmss}}
\def\dyne{{\rm\,dyne}}
     
%\note macro produces sequentially numbered footnotes at bottom of page
%\foot macro produces sequentially numbered footnotes inserted in text
\def\note#1{\footnote{$^{\the\notenumber}$}{#1}\global\advance\notenumber by 1}
\def\foot#1{\raise3pt\hbox{\eightrm \the\notenumber}
     \hfil\par\vskip3pt\hrule\vskip6pt
     \noindent\raise3pt\hbox{\eightrm \the\notenumber}
     #1\par\vskip6pt\hrule\vskip3pt\noindent\global\advance\notenumber by 1}
\def\propo{\propto}
\def\larrow{\leftarrow}
\def\rarrow{\rightarrow}
\def\sectionhead#1{\penalty-200\vskip24pt plus12pt minus6pt
        \centerline{\bbrm#1}\vskip6pt}
     
%\Dt and \dt put Newton's notation dots above upper and lower case chars
\def\Dt{\spose{\raise 1.5ex\hbox{\hskip3pt$\mathchar"201$}}}    % upper case
\def\dt{\spose{\raise 1.0ex\hbox{\hskip2pt$\mathchar"201$}}}    % lower case
\def\llangle{\langle\langle}
\def\rrangle{\rangle\rangle}
\def\ldotss{\ldots}
\def\del{\b\nabla}
     
% equation numbering
%\new macro produces sequentially numbered equations by writing \eqno(\new)
%at end of displayed equations
\def\new{{\rm\chaphead\the\eqnumber}\global\advance\eqnumber by 1}
%to refer to an equation which is 5 equations back, write "equation (\ref5)"
\def\ref#1{\advance\eqnumber by -#1 \chaphead\the\eqnumber
     \advance\eqnumber by #1 }
%\last macro is like \new except counter is not advanced. Useful for equations
%which are in parts a and b.
\def\last{\advance\eqnumber by -1 {\rm\chaphead\the\eqnumber}\advance
     \eqnumber by 1}
%to name an equation, place command "\eqnam{\Poisson}" before equation, and
%thereafter "equation(\Poisson)" will generate the proper equation number.
\def\eqnam#1{\xdef#1{\chaphead\the\eqnumber}}
     
%figure numbering
%\nfig macro assigns number to a figure
\def\nfig{\chaphead\the\fignumber\global\advance\fignumber by 1}
%\nfiga permits a,b,c etc. to be added to figure number
\def\nfiga#1{\chaphead\the\fignumber{#1}\global\advance\fignumber by 1}
\def\rfig#1{\advance\fignumber by -#1 \chaphead\the\fignumber
     \advance\fignumber by #1}
%\def\fignam#1{\xdef#1{\chaphead\the\fignumber}}
%reference macros. To generate reference to a paper in Ap.J. volume 300, p.123
%write \apj{Claus, S. 1990.}{300}{123}
\def\refindent{\par\noindent\parskip=3pt\hangindent=3pc\hangafter=1 }

\def\apj#1#2#3{\refindent#1,  {ApJ,\ }{\bf#2}, #3}
\def\apjsup#1#2#3{\refindent#1,  {ApJS,\ }{\bf#2}, #3}
\def\aasup#1#2#3{\refindent#1,  { A \& AS\ }{\bf#2}, #3}
\def\aas#1#2#3{\refindent#1,  { Bull. Am. Astr. Soc.,\ }{\bf#2}, #3}
\def\apjlett#1#2#3{\refindent#1,  { ApJL,\  }{\bf#2}, #3}
\def\mn#1#2#3{\refindent#1,  { MNRAS,\ }{\bf#2}, #3}
\def\mnras#1#2#3{\refindent#1,  { M.N.R.A.S., }{\bf#2}, #3}
\def\annrev#1#2#3{\refindent#1, { ARA \& A,\ }
{\bf2}, #3}
\def\aj#1#2#3{\refindent#1,  { AJ,\  }{\bf#2}, #3}
\def\phrev#1#2#3{\refindent#1, { Phys. Rev.,}{\bf#2}, #3}
\def\aa#1#2#3{\refindent#1,  { A \& A,\ }{\bf#2}, #3}
\def\nature#1#2#3{\refindent#1,  { Nature,\ }{\bf#2}, #3}
\def\icarus#1#2#3{\refindent#1,  { Icarus, }{\bf#2}, #3}
\def\pasp#1#2#3{\refindent#1,  { PASP,\ }{\bf#2}, #3}
\def\appopt#1#2#3{\refindent#1,  { App. Optics,\  }{\bf#2}, #3}
\def\spie#1#2#3{\refindent#1,  { Proc. of SPIE,\  }{\bf#2}, #3}
\def\opteng#1#2#3{\refindent#1,  { Opt. Eng.,\  }{\bf#2}, #3}
\def\refpaper#1#2#3#4{\refindent#1,  { #2 }{\bf#3}, #4}
\def\refbook#1{\refindent#1}
\def\science#1#2#3{\refindent#1, { Science, }{\bf#2}, #3}
     
\def\chapbegin#1#2{\eject\vskip36pt\par\noindent{\chapheadfont#1\hskip30pt
     #2}\vskip36pt}
\def\sectionbegin#1{\vskip30pt\par\noindent{\bf#1}\par\vskip15pt}
\def\subsectionbegin#1{\vskip20pt\par\noindent{\bf#1}\par\vskip12pt}
\def\topic#1{\vskip5pt\par\noindent{\topicfont#1}\ \ \ \ \ }
     
%\ltsim and \gtsim produce > and < signs with twiddle underneath
\def\ltsim{\mathrel{\spose{\lower 3pt\hbox{$\mathchar"218$}}
     \raise 2.0pt\hbox{$\mathchar"13C$}}}
\def\gtsim{\mathrel{\spose{\lower 3pt\hbox{$\mathchar"218$}}
     \raise 2.0pt\hbox{$\mathchar"13E$}}}
     
%\sec produces arcsec symbol so that 3\sec5 produces 3."5 with the second
%symbol and the period aligned.
\def\sec{\hbox{$^s$\hskip-3pt .}}
\def\gg{\hbox{$>$\hskip-4pt $>$}}
%\hoffset=1.0truein
%\voffset=0.8truein
\parskip=3pt
\def\gapprox{$_ >\atop{^\sim}$}     %Greater than over approximately (wiggle).%
\def\lapprox{$_ <\atop{^\sim}$}     %Less than over approximately.%
\def\apequal{\mathrel{\spose{\lower 1pt\hbox{$\mathchar"218$}}
     \raise 2.0pt\hbox{$\mathchar"218$}}}

\def\oforder{$\sim$} \def\inv{$^{-1}$}
\def\>={$\geq$} \def\<={$\leq$} \def\ks{km s\inv} \def\kms{km s\inv}
\def\lith{$h$} \def\sig{$\sigma$} \def\sigp{$\sigma^{\prime}_r$}
\def\meanz{$\overline \upsilon$} \def\nc{$N_c$} \def\rc{$r_c$}
\def\twidle{$\sim$} \def\sigmar{$\sigma_r$}
\def\Mstar{{M_B}^*}
\def\Mdot{M_{\odot}}

\title{THE PROPERTIES OF POOR GROUPS OF GALAXIES:}  
\title{I. SPECTROSCOPIC SURVEY AND RESULTS} 
\author{Ann I. Zabludoff\altaffilmark{1,2} and John S. Mulchaey\altaffilmark{1}}
\altaffiltext{1}{Observatories of the
Carnegie Institution of Washington, 813 Santa Barbara St., Pasadena,
CA 91101, E-mail: mulchaey@pegasus.ociw.edu}
\altaffiltext{2}{UCO/Lick Observatory and Board of Astronomy and
Astrophysics, University of California at Santa Cruz, Santa Cruz, CA,
95064, E-mail: aiz@ucolick.org} 

\bigskip
\centerline {Accepted for publication in {\it The Astrophysical Journal}}

\singlespace
\abstract{
We use multi-fiber spectroscopy of 12 poor groups of
galaxies to address:  (1) whether the groups are bound
systems or chance projections of galaxies along
the line-of-sight, (2) why the members of each group have not already
merged to form a single galaxy, despite the groups' high galaxy densities, 
short crossing times, and likely environments
for galaxy-galaxy mergers,
and (3) how galaxies might evolve in these groups, where the collisional
effects of the intra-group gas and the tidal influences
of the global potential are
weaker than in rich clusters.  Each of the 12 groups
has fewer than $\sim$ five cataloged members in the literature.  
Our sample consists of 1002 galaxy velocities, 
280 of which are group members.
The groups have mean recessional velocities between 1600 and 7600 \ks.  Nine 
groups, including three Hickson compact groups, have the extended
X-ray emission characteristic of an intra-group medium
(Mulchaey \& Zabludoff 1997 (Paper II)).

We conclude the following:

{\it (a) The nine poor groups
with diffuse X-ray emission are bound systems
with at least $\sim $20-50 group
members to $M_B \sim -14$ to $-16 + 5$log$_{10}$ \lith}.
The large number of group members, the significant early-type population
(up to $\sim 55\%$ of the membership) and its concentration
in the group center,
and the correspondence of the central, giant elliptical with
the optical and X-ray group centroids argue that
the X-ray groups are not radial superpositions
of unbound galaxies.  The velocity dispersions of the X-ray groups range from
190 to 460 \ks.  We are unable to determine if
the three non-X-ray groups, which have lower
velocity dispersions ($< 130$ \ks) 
and early-type fractions ($= 0$\%), are also bound.

{\it (b) Galaxies in each X-ray-detected 
group have not all merged together, because
a significant fraction of the group mass lies outside of the galaxies and
in a common halo.}
The velocity dispersion of the combined group sample
is constant as a function of radius out to the virial 
radius of the system (typically $\sim 0.5$\lith\inv\ Mpc).
The virial mass of each group ($\sim 0.5$-$1\times 10^{14}h^{-1} \Mdot$)
is large compared with the mass in the X-ray gas and in the galaxies 
({\it e.g.}, $\sim 1 \times 10^{12} h^{-5/2} \Mdot$ 
and $\sim 1 \times 10^{13} h^{-1} \ \Mdot$, 
respectively, in NGC 533).  These results imply that
most of the group mass is in a common, extended halo.
The small fraction ($\sim 10$-$20\%$) of group mass 
associated with individual galaxies suggests that the rate of galaxy-galaxy
interactions is lower than for a galaxy-dominated system
(Governato \etal 1991; Bode \etal 1993; 
Athanassoula 1997), allowing these groups to
virialize before all of their galaxies merge and
to survive for more than a few crossing times.  

{\it (c) The position of the giant, brightest elliptical in each X-ray group
is indistinguishable from the center of the group potential, as defined by
the mean velocity and the projected spatial centroid of the group galaxies.}
This result suggests that dominant cluster ellipticals,
such as ``cD" galaxies (Matthews, Morgan, \& Schmidt 1965), may form
via the merging of galaxies in the centers of 
poor group-like environments.
Groups with a central, dominant elliptical may
then fall into richer clusters (Merritt 1985).  This scenario explains
why ``cD"s do not always lie in the spatial and kinematic
center of rich clusters (Zabludoff \etal 1990; 
Dunn 1991; Zabludoff \etal 1993), 
but instead occupy the centers of
subclusters in non-virialized clusters (Geller \& Beers 1983; Bird 1994;
Beers \etal 1995).

{\it (d) The fraction of early-type galaxies in our poor groups varies 
significantly, ranging from that characteristic of the field
($\simless 25$\%) to that of
rich clusters ($\sim 55$\%).}
The high early type fractions are particularly surprising, because
all of the groups in this sample have substantially
lower velocity dispersions (a factor of $\sim 2$-5)
and galaxy number densities (a factor of $\sim 5$-20)
than are typical of 
rich clusters.  Hence, the effects of disruptive mechanisms like
galaxy harassment (Moore \etal 1996)
on the morphology of poor group galaxies are weaker
than in cluster environments.  
In contrast, the kinematics of poor groups make them preferred sites for
galaxy-galaxy mergers (Barnes 1985; Aarseth \& Fall 1980;
Merritt 1985), which may alter
the morphologies and star formation histories of some group members.
If galaxy-galaxy interactions 
are not responsible for the high early type fractions, it is possible that the
effects of environment are relatively unimportant at the current epoch and that
the similarity of the galaxy populations
of rich clusters and some poor groups reflects
conditions at the time of galaxy formation.

{\it (e) The fraction of early-type group members that have experienced
star formation within the last $\sim$ 2\lith\inv\ Gyr
is consistent with that in rich clusters
with significant substructure ($\sim 15\%$; Caldwell \& Rose 1997).}
If some of the subclusters in these rich, complex
clusters are groups that have
recently fallen into the cluster environment,
the similarity between the star formation histories of the
early types in the subclusters and of those in our
sample of field groups indicates that the cluster environment
and associated mechanisms like ram pressure stripping (Gunn \& Gott 1972) are
not required to enhance and/or quench
star formation in these particular galaxies.  
If the recent star formation is tied to the
external environment of the galaxies and not to internal instabilities,
it is more likely that galaxy-galaxy encounters have altered
the star formation histories of some early type galaxies in groups
and in subclusters.

\bigskip
\bigskip
\noindent{\it Subject headings}:  galaxies: clustering ---
galaxies: distances and redshifts ---
galaxies: elliptical and lenticular, cD ---
galaxies: evolution ---
galaxies: interactions ---
cosmology: dark matter ---
cosmology: large-scale structure of Universe
}

\vfill\eject
\section{Introduction}
Most galaxies in the local universe, including our own Galaxy,
belong to poor groups of
galaxies.  Despite the ubiquity of the group environment, we know little
about the matter content of groups and the evolution of group galaxies
outside of the Local Group.
Because poor groups typically contain fewer than five bright ($\simless M^*$)
galaxies, studies to date have been hampered by small number statistics.
Some of the critical, unanswered questions are
(1) whether poor groups are in fact bound
systems with significant populations of fainter members, (2)
why many poor groups, with their high galaxy densities, 
short crossing times, and favorable environments
for galaxy-galaxy mergers, survive long enough to be cataloged,
and (3) how galaxies might evolve in an environment where
the influences of the intra-group medium and the global potential are
weak compared with those in rich clusters.
The advent of multi-object spectroscopy now makes it possible to address these
questions in unprecedented detail.  In this paper, we present the
first results from a fiber spectroscopic survey
of 12 poor groups of galaxies.

The issue of whether many poor groups, even
those identified from redshift surveys, are bound systems
instead of chance superpositions of galaxies along the line-of-sight
has been a puzzle.  The existence of one poor group, our Local Group,
is unchallenged.  In contrast,
Ramella \etal (1989) show that $\sim 30\%$ of groups of three or four
galaxies in the CfA Redshift Survey (Huchra \etal 1995) 
are probably unbound, geometric
projections.  One useful approach in finding bound systems is to
identify those, like certain Hickson
compact groups, with apparently interacting 
members (Rose 1977; Hickson 1982; de Oliveira \& Hickson 1994). 
Yet these systems are also subject to projection effects
({\it i.e.,} a pair of interacting galaxies with two interlopers; Mamon 1992)
and, because they constitute only a small fraction of cataloged groups,
may not be dynamically representative.
Another strategy is to search for 
poor groups with diffuse X-ray emission
(Mulchaey \etal 1993; Ponman \& Bertram 1993; Pildis \etal 1995), 
in which the existence of a
common gravitational potential is suggested by the intra-group gas
({\it e.g.,} Ostriker \etal 1995).  
ROSAT images reveal that at least
$25\%$ (22 of 85) of Hickson compact groups (Ponman \etal 1996)
have such an intra-group medium.  
However, 
the potentials of some poor groups may be too shallow for
emission from hydrostatic gas to reach 
detectable levels (Mulchaey \etal 1996a).
It is even possible that the gas, like the galaxies in some cases,
is merely a
projection of unbound material in a filament alone the line-of-sight
(Hernquist \etal 1995).
To determine whether a poor group is a bound system and to explore its
kinematics in detail, we must spectroscopically identify more members.
Furthermore, if fainter populations of galaxies do exist in
these groups, then we will have the statistics necessary to quantify
the mass associated with
the galaxy, X-ray gas, and dark matter components and to
better understand how galaxies may evolve in such environments.

If some poor groups are bound systems, then another critical question is
why they exist at all.  Poor groups have
higher galaxy densities than the field 
and lower velocity dispersions than cluster cores,
making them favorable sites for galaxy-galaxy 
mergers (Barnes 1985).  Galaxies are tidally interacting or merging
in many Hickson compact groups
(de Oliveira \etal 1994; Longo \etal 1994; Hunsberger \etal 1996; 
Yun \etal 1997).
The likelihood of mergers and the short group crossing times ($\simless 
0.05$ of a Hubble time)
suggest that most groups should have already merged into one object.  
Therefore, either bound groups are collapsing for the
first time or only a small fraction of the group mass
is tied to the galaxies, lowering the rate of 
galaxy-galaxy interactions relative to
a galaxy-dominated system and allowing the group to
survive many crossing times (Governato \etal 1991; 
Bode 1993; Athanassoula 1997).
To resolve this issue by measuring the
underlying mass distribution of poor groups, 
we need to improve the statistics of group
membership with an extensive spectroscopic survey.

Once we know if a group is real and how much mass is associated with
its galaxies, we can investigate the influences
of group environment on galaxy evolution.  For example,
because ``cD" galaxies in clusters 
lie in regions of high local density
(Beers \& Geller 1983; Zabludoff \etal 1990; Beers \etal 1995), 
but not always in 
the center of the global cluster potential,
these galaxies may evolve first in poor group-like environments 
prior to the
final collapse and virialization of the cluster as a whole
(Merritt 1985).  If so, then ``cD"s are likely to form
via galaxy-galaxy mergers in the center of a collapsing group,
where the conditions for mergers are most favorable (Merritt 1984;
Tremaine 1990).
Over the lifetime of the group, dynamical friction or radial orbits
may bring in more galaxies to merge with the ``cD."
X-ray-detected poor groups, which almost always contain
a giant ($\simless \Mstar - 1$)
elliptical near the peak of the X-ray emission,
are ideal laboratories for testing this picture of ``cD" formation.
If the group environment is the birthplace of ``cD"s, then
the giant elliptical will be coincident with
the centroid of the projected spatial distribution of galaxies
and will have little peculiar velocity with respect to the mean of
the system.  An extensive spectroscopic survey
of group members will allow us to test this formation hypothesis directly
in the poor group environment.

The factors that might affect the evolution of galaxies in poor groups
are different from those present in rich clusters.
Some of the proposed cluster-based processes,
such as ram pressure stripping (Gunn \& Gott 1972) and 
galaxy harassment (Moore \etal 1996),
are less effective
in group environments, where 
the number density of bright galaxies and the global velocity
dispersion are
small compared with clusters. 
The lower velocity dispersions of poor groups suggest instead that
galaxy-galaxy
interactions ({\it e.g.}, close tidal encounters
or mergers) are likely to dominate
any environmentally-dependent evolution of galaxies in groups.
If clusters evolve hierarchically by accreting
poor groups of galaxies (subclusters), members of an infalling group
have recently experienced 
the hot, dense cluster environment for the first time.
Therefore, galaxies in poor groups in the field
are a control sample for understanding the factors that
influence the evolution of their counterparts in subclusters.
For example, we can compare the
morphologies and recent star formation histories of
galaxies in the substructures of complex clusters like Coma
(Caldwell \& Rose 1997) with those of galaxies in
poor field groups.  Differences between the samples would argue
that cluster environment is important in transforming galaxies
at the present epoch.
On the other hand, the lack of such differences
would suggest, as the simplest explanation,
either that star formation and morphology are influenced by
mechanisms present in both field groups and subclusters, such as galaxy-galaxy
encounters, or that the effects of environment on
galaxies are insignificant compared
with conditions at the time of galaxy formation.

This paper is organized in four sections 
that address whether poor groups are bound systems, 
why some groups have not yet disappeared, 
and how the group environment may affect galaxy evolution.
Section $\S2$ describes the
the sample of 12 poor groups, the fiber spectra, and the classification of
group members as early or late type.  We discuss our results
in $\S3$, including the membership and global velocity dispersion
of each group,
the composite group velocity dispersion profile, 
the location of the dominant elliptical relative to the kinematic
and projected spatial
centers of groups, the fraction of early type galaxies
in groups, and the fraction of these early types that show evidence for
recent or on-going star formation.
Section 4 summarizes our conclusions.

\section{The Data}

\subsection{The Group Sample}

A poor group is defined as an apparent system of fewer than five bright
($\simless \Mstar$) galaxies.
Poor groups in the literature have been identified optically and 
fall into several classes that can be distinguished by
their X-ray properties and bright galaxy morphologies (Mulchaey \etal 1996b).  
Groups with detectable intra-group gas typically have a giant
($\simless \Mstar - 1$) elliptical that is the brightest
group galaxy (BGG) and
that lies near or on the peak of the X-ray emission
(Mulchaey \etal 1996b).  In contrast, groups without extended
X-ray emission tend to optically
resemble the Local Group, which consists of a
few bright late-type galaxies and their satellites.
If some non-X-ray-detected, late-type-dominated groups evolve
via such mechanisms as galaxy mergers, dynamical friction,
gas stripping from galaxies, and/or infalling primordial gas,
into groups with an central, giant elliptical and a detectable intra-group
medium,
then groups in transition may form a third class of objects.
We would expect the cores of such systems to have signatures of recent
dynamical evolution, including interacting galaxies and
X-ray gas that does not coincide with the galaxies.
Existing group catalogs, such as the Hickson survey of poor, compact groups
(Hickson 1982), contain all three classes of poor groups
({\it e.g.,} de Oliveira \& Hickson 1994; de Carvalho \etal 1994).
To construct our sample, we select poor
groups from these three classes
that have complementary ROSAT Position Sensitive Proportional Counter
(PSPC) X-ray images.

The sample consists of 12 nearby ($1500 < cz <  8000$ \ks), 
optically-selected groups from the
literature (NASA/IPAC Extragalactic Database (NED), Helou \etal 1991) 
for which there are existing, sometimes serendipitous, pointed PSPC
observations of the fields in which the groups lie.
The integration times of the PSPC images are sufficiently
long ($> 2000$ sec; Mulchaey \& Zabludoff 1997 (Paper II)) to allow the
detection of all groups with X-ray luminosities
of $L_X \simgreat 10^{40} h^{-2}$ erg s\inv\ within this range of
radial velocities.  These groups are chosen to be accessible from
Las Campanas Observatory and to have at least
three bright galaxies with known
redshifts and magnitudes.  

Of the 12 groups, we observed five in August 1995 and seven
in April 1996 with the
multi-fiber spectrograph (Shectman \etal 1992) and 2D-Frutti detector
mounted on the du Pont 2.5m telescope at the Las Campanas Observatory.
The selection of groups for the first spectroscopic run differed from
the second.  Before the first run, we did not know which
of the five groups had detectable, extended X-ray emission.
We chose two
groups with the central, giant elliptical characteristic of systems
with intra-group media (Mulchaey \etal 1996b):  NGC 533 and NGC 741.
For contrast, we selected three other groups, NGC 491, NGC 664, and
NGC 7582, that have optical morphologies more
akin to the Local Group.  
(NGC 664 is a previously uncataloged group, identified as
four galaxies at the same redshift in the NED database,
whose position on the sky, mean radial velocity, high Galactic
latitude, and serendipitous
ROSAT pointing made it suitable for the observing program.)
Of the seven groups observed during the second run, all were known
X-ray detections, including
HCG 90, a possible transitional object with
several interacting galaxies in its core (Longo \etal 1994)
and a marginal X-ray detection (Ponman \etal 1996).  Therefore, 
our group sample is not representative
of published group catalogs, in which less than half of the groups
are X-ray-detected.  Instead, the sample is weighted toward
X-ray groups, because of the likelihood that they are
bound systems and the most evolved poor groups.  

In the final reduction of the PSPC data, we detect extended emission in
nine of the 12 groups:  NGC 533, NGC 741, NGC 2563, NGC 4325, NGC 5129,
NGC 5846, HCG 42, HCG 62, and HCG 90.
The X-ray gas in HCG 90 is
asymmetric and not coincident with the galaxies, suggesting that
this group is in fact dynamically evolving.
The three Local Group-like targets, NGC 664, 
NGC 491, and NGC 7582, are not detected.  Paper II 
includes a full discussion of the analyses of the  
X-ray and STScI/Digitized Sky Survey images.

To obtain galaxy targets in each group field
over the $1.5\times 1.5$ degree field of the
fiber spectrograph, we used coordinates,
star/galaxy classifications, and magnitudes from the
Automated Photographic Measuring survey (APM, Maddox \etal 1990).
The uncalibrated, relative magnitudes
drawn from the blue plate scans were
sufficient to identify the $\sim 150$ brightest
galaxies in each field.  
Using the STScI/Digitized Sky Survey,
we checked all potential targets typed as galaxies by eye,
discarding stars and plate flaws.  Inspection of the
DSS revealed that about $\sim$5-10$\%$ of the brightest galaxies
were not included in the APM catalog,
so we added the most obvious omissions to our final target lists.

To quantify the sample incompleteness due to the
underrepresentation of bright galaxies in the APM catalog, we ran the 
FOCAS program (Jarvis \& Tyson 1981) 
on the STScI/Digitized Sky Survey scan for each group
after the completion of both observing runs.
We used the FOCAS classification scheme to 
identify the $\sim 150$ brightest galaxies in each field.
We visually checked all objects typed as galaxies
by FOCAS, eliminating
stars and plate flaws.  Inspection of the scans
confirmed that that the FOCAS catalogs were more complete
than the APM output, so we adopted the FOCAS catalogs
as the master lists from which we subsequently determined the
completeness of the group samples ($\S3.1$).  
We matched the FOCAS coordinates to the APM coordinates and
list the former in Table 1.

\subsection{The Spectra}

We obtain fiber spectra for $\sim 50$-100 of the brightest 
galaxies in each of the 12 group fields.  In total,
we measure 963 galaxy spectra.
Each spectrum is extracted from the two-dimensional array, flat-fielded, 
wavelength-calibrated, and finally sky-subtracted based on
the flux normalization of the 5577 \AA, 
5890 \AA, and 6300 \AA\ night sky lines.  The spectra
have a resolution of $\sim$5-6 \AA, a pixel scale of $\sim 3$ \AA, and a
wavelength range of 3500-6500 \AA.
The average signal-to-noise
$S/N$ in the continuum around the H$\beta \ \lambda 4861$, 
H$\gamma \ \lambda 4340$, and H$\delta \ \lambda 4102$
absorption lines is typically $\sim 8$
(calculated by determining the ratio of the mean square deviation about the
continuum near each absorption line to the mean continuum
at the absorption line, after excluding the absorption line
and any nearby sky lines).
The 3.5 arcsec fiber aperture subtends
projected physical diameters between 0.3\lith\inv\ kpc
and 1.2\lith\inv\ kpc at the distances of the groups
(from 16 to 72\lith\inv\ Mpc; $q_0 = 0.5$ and $H_0 =
100h$ \ks\ Mpc\inv\ are used throughout this paper).
Hence, the fibers only sample
light from the core of each group member.

We determine the radial velocities using the cross-correlation routine
XCSAO and the emission line finding routine EMSAO in the RVSAO
package in IRAF (Mink \& Wyatt 1995).  The velocities in Table 1 are either
emission line velocities, absorption line velocities, or a weighted
average of the two
(see Shectman \etal 1997 (their $\S2.2$) or Lin 1995 
for a discussion of the cross-correlation
templates and the spectral lines typically observed).  
We compute velocity 
corrections to the heliocentric reference frame with the
IRAF/HELIO program.
By adding 39 galaxy velocities from NED, we obtain
a total of 1002 velocities in the group fields.
For galaxies projected within 0.3\lith\inv\ Mpc of the group
center and brighter than an
absolute magnitude of $M_B \sim -17 + 5$log$_{10}$ \lith, 
the samples range in completeness from $\sim 60$ to $100\%$
of the FOCAS catalog (Figure 4), although we observe
galaxies out to projected radii as large as 0.95\lith\inv\ Mpc
and to limits as faint as
$M_B \sim -14$ to $-16 + 5$log$_{10}$ \lith, 
depending on the distance to the group. 

We estimate the velocity zero-point correction
and external velocity error
by comparing our velocities with HI velocities from NED.  Figure 1 shows
the residual for 39 galaxies as a function of our internal velocity
error estimate.  We use only those HI velocities with
quoted errors of $< 30$ \ks.
The mean residual of 13 \ks\ (solid line) is small
compared with the {\it rms} deviation of the residuals ($\sim 80$ \ks) and is
consistent with
the mean velocity of 94 stars ($-17$ \ks) 
that were serendipitously observed with the
same instrument (dashed line).  Therefore, we do not apply a zero-point
correction to the velocities.

We adopt the {\it rms} deviation of the residuals (80 \ks),
which is constant over the range of internal errors, as
the true velocity error when the internal or NED error
is smaller than 80 \ks; otherwise, we list the internal or NED error.
Our error estimates are consistent with
the average external error estimate of $\sim$70 \ks\ for 
the Las Campanas Redshift Survey (Shectman \etal 1996), which employs the same
fiber spectrograph setup.

\subsection{Galaxy Classification}

We morphologically classify
the group members (as defined in $\S3.1$) 
that have apparent magnitudes of $m_B \sim 17$ or brighter.
This apparent magnitude limit is equivalent to absolute
magnitude limits ranging from 
$M_B \sim -14$ to $-17 + 5$log$_{10}$ \lith, 
depending on the distance to the group.
The galaxies are typed as either
early (E and S0) or late (all others) from R-band CCD images
or from STScI/Digitized Sky Survey POSS E or SERC $B_J$ images.
The CCD and the scanned plate data overlap for
$60\%$ of the 188 typed galaxies, and our classifications
from these two media agree.  Because we use
only two broad classifications, there is complete
agreement between the independent classifications of the authors.
Our classifications are also $\sim 85\%$ consistent with
classifications in NED for the 100 group galaxies with
published morphologies.
Most of the discrepancies arise when
our classification is S0/a and NED's type is S0,
because we define S0/a transitional galaxies as late types.

Table 1 lists the galaxy name from our catalog, 
J2000 coordinates from FOCAS
using the DSS/STScI plate solution, heliocentric radial velocity ($\upsilon$),
type of velocity measurement
(from absorption lines ``0", emission lines ``1", or a 
combination of both ``2"), 
morphological classification 
(either early ``e" or late ``s" type), and the medium (plate ``p" or
CCD ``c") that we use to classify the galaxy.  We note
the 39 heliocentric velocities obtained from NED and the
references therein with an ``N" in the last column.  
An example of the format of Table 1 is given in the text.
The full version of the table that includes the 1002 galaxy velocities
in the sample is available on
the CDROM accompanying this volume.

\section{Results and Discussion}

\subsection{Group Membership}

Are poor groups bound systems or chance superpositions of galaxies along
the line-of-sight?
Our spectroscopic survey samples
the fields of groups whose membership previously totaled
less than five bright ($\simless \Mstar$) galaxies.
The detection of a significant population of fainter members would be a
first step in demonstrating that these poor groups
are not just geometric projections of unbound galaxies.

We determine the
galaxy membership of each group from a pessimistic, $3\sigma$-sampling 
algorithm (Yahil \& Vidal 1977).  We use the
statistical bi-weight estimators of location (mean velocity)
and scale (velocity dispersion) to identify $3\sigma$ outliers 
in the distribution of galaxy velocities within $\pm 3000$ \ks\ of the
center of the main peak (see Beers \etal 1990 for a description of the
bi-weight estimators).
On each successive iteration, the $3\sigma$ outliers are removed and the
location and scale of the peak re-calculated.  We halt
the procedure prior to the removal of the last set of outliers.

The resulting membership of each of the 12 groups 
is indicated by the shaded histograms in
Figure 2, which shows the galaxy velocity distributions from
0 to 30000 \ks.  The width of the velocity bins is 250 \ks,
roughly $3\times$ the typical external error.  In total, there are
280 group members.
The first nine groups are detected by the ROSAT
PSPC, the last three are not (see Paper II).
The histograms of the X-ray groups reveal large
populations of group members down to absolute magnitudes 
of $M_B \sim -14$ to $-16 + 5$log$_{10}$ \lith.
Because the membership algorithm can not be applied to
the small number of galaxies in the peak of each
non-X-ray group, we accept all the galaxies within contiguous bins
as group members.  As a result, the membership
for the non-X-ray groups may be overestimated.   

We show the projected spatial distributions of the group members in Figure 3.
The angular size of each plot
is $1.62\times 1.62$ degrees
(the fiber spectrograph field is
$1.5\times 1.5$ degrees).  Each tickmark corresponds to 5.7 arcmin.
Digitized scans of Palomar Sky Survey
or UK Schmidt plates from the STScI/DSS
are not complete for every group.
Hence, we mark the boundary of the unsampled regions with a dashed line:
the westernmost fifth of
the NGC 741 field, in the northernmost fifth of the NGC 5129 field,
in the northernmost fifth of the HCG 90 field, and the northernmost tenth
of the NGC 7582 field.  The morphological types of galaxies with apparent
magnitudes of $m_B \sim 17$ or brighter are indicated by ``0" for early
and ``S" for late.  The filled circles mark the
untyped group members.  The scale bar below each group name
is 0.3\lith\inv\ Mpc.  In the X-ray groups, as in rich clusters,
the early types concentrate more in the group centers than do
the late types.
There are no early types in the
three non-X-ray groups (see $\S3.5$).

The number counts of group members in the velocity histograms in Figure 2
are not directly comparable, because each group field
is sampled over a different
physical radius and to a different absolute magnitude.
To compare the galaxy number densities of the groups, we
subsample each system within a projected radius of 0.3\lith\inv\ Mpc and
to an absolute magnitude of $M_B \sim -17 + 5$log$_{10}$ \lith. 
Figure 4 shows the 
observed number counts of group members within these limits (shaded).
To roughly compensate for incomplete sampling down to the magnitude
limit (completeness is indicated
by the fraction above each histogram bar), we assume that
the fraction of all unobserved galaxies that are group members 
is the same as the fraction of all observed galaxies that are members.
The white histogram shows these ``corrected" group galaxy counts.
Our rough calibration of the FOCAS magnitudes introduces more uncertainty into
the ``corrected" galaxy counts, so a
small difference between the counts of two groups is not significant.
Also note that the ``corrected" counts are lower limits
for two groups, NGC 7582 and NGC 5846, for which the $1.5\times 1.5$
degree fiber field corresponds to less than 0.3\lith\inv\ Mpc 
(0.21 and 0.24\lith\inv Mpc, respectively).
Nevertheless, as this plot shows, non-X-ray-detected groups
have lower galaxy densities than are typical of X-ray groups.
For the same radial and magnitude cuts, the core of the Coma cluster (NED)
has 83 galaxies (a lower limit because we make no correction
for incompleteness).  Thus, the galaxy density of the Coma
core is $\sim$5-$20\times$ that of the poor group cores.

By sampling to deeper magnitudes and to larger radii than past studies, 
we find that the physical extents of the poor groups in Figure 3, even those
of HCG 42, HCG 62, and HCG 90, are larger than the typical
values in the literature for Hickson compact groups (Hickson \etal 1992).
The number densities of galaxies
within 0.3\lith\inv\ Mpc and with $M_B \simless -17 + 5$log$_{10}$ \lith\
in HCG 42, HCG 62, and HCG 90 
(Figure 4) are a factor of $\sim 70$-700 times 
lower than those inferred from the brightest four members (Hickson, 
Kindl, \& Huchra 1988).
These discrepancies result from the method of selecting Hickson
groups, which are defined as concentrations of four or five
bright galaxies within a projected radius of $\simless 0.1$\lith\inv\ Mpc
(see also de Carvalho \etal 1994).  
The galaxy number densities of the Hickson compact 
groups are not significantly different from those of the other poor groups
in our sample.  In general, the presence or absence of diffuse X-ray emission
better differentiates between poor groups of high and
low galaxy densities (and
velocity dispersions ($\S3.2$) and early type fractions ($\S3.5$)).

Could the X-ray groups be superpositions of three or
four unbound, field galaxies and their fainter satellites?
This explanation is unlikely, because the central, giant elliptical
is not typical of field galaxies
and the early type galaxies tend to concentrate in the group core.
In addition, studies of the satellite populations of 
$\simless \Mstar$ field galaxies with
the same fiber spectrograph setup (Zaritsky \etal 1997),
within similar projected radii from the primary galaxy, and to comparable
absolute magnitude limits as our group sample
find on average one satellite for each primary.
The ratios of faint to giant galaxies in the X-ray groups exceed 
that expected from satellite statistics alone.

The large populations of fainter galaxies 
and the central concentration of early types in the X-ray groups
suggest that these groups are bound systems.
The dynamical state of the non-X-ray groups is less obvious.
In $\S3.2$, 3.3, and 3.4, we use the global kinematic properties,
the velocity dispersion profile, and 
the position of the central, giant elliptical relative to the
kinematic and projected spatial group centroids
as additional tests of whether the X-ray groups are bound and possibly
virialized.

\subsection{Group Velocity Dispersions}

\subsubsection{Results}
To date, the line-of-sight velocity dispersions (\sigmar) of poor groups have
been uncertain due to their determination from only a few galaxies.
This uncertainty has translated into a large uncertainty in cosmologically
important properties like the underlying mass distribution
and the global baryon fraction.
Because we increase the membership by a factor of 10 in many of our groups, 
we can, for the first time, determine poor group
l-o-s velocity dispersions with sufficient precision ($\simless 20\%$) that the
differences among them are statistically meaningful.
The calculation of the mean velocity \meanz\ and 
\sigmar\ for each group is based on the bi-weight 
estimators of
location and scale (Beers \etal 1990) corrected for cosmological
effects.  These estimators are more robust
than the standard mean and velocity dispersion
({\it e.g.}, Danese \etal 1980), but, for this sample, the
standard velocity
dispersion is always within the $68\%$ confidence limits of \sigmar.
The velocity dispersions of the X-ray groups range over more
than a factor of two, from
190 \ks\ in HCG 90 and 210 \ks\ in HCG 42 to 430 \ks\ in NGC 741 
and 460 \ks\ in NGC 533.

The results of $\S3.3$ argue that the group velocity dispersion
remains fairly constant with radius out to at least 0.5\lith\inv\ Mpc.
However, to confirm that the range of velocity dispersions
above does not result from the different physical radius to
which each group is sampled by the fixed angular size of
spectrograph field, we determine
the membership and the
velocity dispersions within 
a 0.3\lith\inv\ Mpc radius.  The resulting \sigmar 's
are indistinguishable within the $68\%$ confidence 
errors from those determined
from all the data.  Therefore, we use 
the \meanz\ and \sigmar\ determination from 
all the group members in subsequent analyses.

Table 2 lists the group, the group optical projected centroid in J2000
(unweighted by luminosity), 
the total number of galaxies with
measured redshifts in the fiber field
$N_{tot}$, the number of group members $N_{grp}$, the bi-weight estimators
of the mean heliocentric velocity \meanz\ and the line-of-sight 
velocity dispersion \sigmar, and
the physical radius of the group sampled by the fiber field $r_{samp}$. 
For the groups where $r_{samp} \geq 0.67$ (the median
pairwise radius $r_p$ for CfA Redshift Survey groups (Ramella \etal 1989)), 
we tabulate $r_p$,
the mean harmonic (virial) radius $r_h$, the 
virial mass $M_{vir}$,
and the ratio of the crossing time
to a Hubble time $t_c/t_H$.  These last four kinematic quantities are
calculated as in Ramella \etal (1989).  
For the five groups with calculated masses
and crossing times, $r_{samp}$ is $\simgreat 20\%$ larger than $r_p$,
suggesting that the limited size of
the fiber field does not artificially reduce our estimate 
of the group extent and
bias the kinematic determinations.

The median value of the harmonic radius of the four X-ray groups is 
0.41\lith\inv\ Mpc,
consistent with the
median $r_h = 0.5$\lith\inv\ Mpc 
for groups in the CfA Redshift Survey (Ramella \etal 1989).
However, $r_h$ depends on how the distribution of
galaxies is biased with respect
to that of the dark matter in the system.
To check that the harmonic radius is an accurate estimate of the virial radius,
we calculate the radius $r_{500}$ that corresponds to
an overdensity of 500, where N-body simulations show that the
dynamical equilibrium hypothesis is satisfied (Evrard, 
Metzler, \& Navarro 1996).
For a poor group with an X-ray temperature typical of our X-ray groups 
(about 1 keV, Paper II),
$ r_{500} = 1.24 \sqrt{(kT / 10 \ {\rm keV}}) \ h^{-1} \ {\rm Mpc} 
= 0.4 \ h^{-1} \ {\rm Mpc} $.  Therefore, we conclude
that biasing between the galaxies and dark matter
does not significantly affect the estimates of the kinematic
quantities in Table 2.

The velocity dispersions of poor groups
in the literature are usually estimated 
from the four or five brightest members.  With our deeper redshift samples,
we can determine the accuracy of these past estimates.
The velocity dispersions calculated from only 
the five brightest galaxies in each group $\sigma_5$
tend to underestimate \sigmar, because the tails of the 
galaxy velocity distribution are not well-sampled.  In the worst
case in our sample, 
the NGC 2563 group, $\sigma_5$ underestimates \sigmar\ by a factor of three.
In total, $\sigma_5$ underestimates \sigmar\ by more than
a factor of 1.5 for five of the nine X-ray groups.
The difference between $\sigma_5$ and \sigmar\ for
the non-X-ray groups is small, because we calculate \sigmar\ itself 
from only $\sim 5$ galaxies.

\subsubsection{Implications}
The $\sim 20$-50 group members, the central concentration of 
early type galaxies, and the short crossing times 
($\simless 0.05$ of a Hubble time) of the X-ray groups
suggest that they are bound systems, not geometric superpositions
of galaxies, and that the group cores are close to virialization or
virialized.  Because we do not detect diffuse X-ray emission or
a significant fainter population in the three non-X-ray groups,
we are unable to determine their dynamical state.
The non-X-ray groups, which
consist of one or two $L^*$ or brighter spirals
with several fainter galaxies that may be satellites,
are morphologically akin to the Local Group
(although our samples are not sufficiently deep to ascertain whether
any group has a dwarf spheroidal population like that of
the Local Group; van den Bergh 1992).
If the non-X-ray groups are dynamically similar to the Local Group,
they are bound (see Zaritsky 1994). 
Our current data do not exclude this possibility --- 
the velocity dispersions of the non-X-ray groups are consistent
with the upper limits on their X-ray luminosities (Paper II).
If the non-X-ray groups are bound, but are collapsing for the
first time like the Local Group (cf. Zaritsky 1994), 
the virial mass of systems like 
NGC 664 in Table 2 is uncertain by a factor of two (as the system is not
yet virialized).  If the non-X-ray groups are just chance superpositions,
their kinematic quantities in Table 2 do not represent the properties
of bound groups.

\subsection{Velocity Dispersion Profile}

\subsubsection{Results}
The ratio of the mass associated with group galaxies to the mass
associated with the common group halo determines the
timescale of galaxy-galaxy interactions and thus the group's
ability to survive for more than a few crossing times 
(Governato \etal 1991; Bode 1993; Athanassoula 1997).
The statistics of our sample make analyzing group kinematics
on an individual basis difficult.  To obtain an understanding of the
underlying mass distribution in poor groups, we combine the
velocity and projected spatial distributions of all of the X-ray group
members.

Figure 5 shows the velocity offset 
vs. the projected radial offset of 204 X-ray group
members from the central, giant elliptical
(the brightest group galaxy, 
hereafter BGG).  The velocity difference
is normalized with the internal 
velocity dispersion of the BGG ($\sigma_{BGG}$).
We include only the seven X-ray groups for which $\sigma_{BGG}$ 
is known (HCG 42, HCG 90, NGC 2563, NGC 5129, NGC 5846, NGC 533, NGC 741; 
McElroy 1995; Trager 1997).
We calculate the normalized velocity dispersion $\delta$
(i.e., the {\it rms} deviation in the normalized velocity offset) within each
radial bin (dashed lines).  

The velocity dispersion of the combined group sample does
not decrease significantly with radius from the central
$\sim 0.1$\lith\inv\ Mpc to at least $\sim 0.5$\lith\inv\ Mpc,
in contrast to the more than 
factor of two decrease in $\delta$ that would be observed if the entire
mass were concentrated within the first bin.
The extended mass is either in the galaxies, in a common
halo through which the galaxies move, or in both the galaxies and a diffuse
halo.  If all the mass were tied to the galaxies,
most of the mass would be associated with the bright, central elliptical
in those groups in which the BGG dominates the light.  For groups with
a few galaxies that have luminosities comparable to the BGG,
$\delta$ would be increased at large radii by subgroups consisting of a
massive galaxy and the subgroup members that are orbiting it.
If this picture were accurate, we would expect the $\delta$ profiles
of groups with several comparably bright galaxies to be shallower than those in
which the BGG is dominant.
However, the combined velocity
dispersion profile of a subsample of two groups (HCG 42 and NGC 741)
in which the BGG dominates the light ({\it i.e.},
the BGG luminosity exceeds the combined luminosity of
the other $L^*$ or brighter galaxies) is indistinguishable from that of
the entire sample.
Therefore, we conclude that most of the group mass lies in a smooth,
extended dark halo.

Is this halo associated with the group as a whole or with the
BGG?  If a galaxy forms inside a dark halo through
dissipative collapse (cf. Blumenthal \etal 1986), baryons concentrate
in the center, deepening the gravitational well and increasing
the velocity dispersion of the galaxy relative to that of the halo.
In all the radial bins in Figure 5, $\delta > 1$, indicating that 
the BGG is on average
dynamically {\it cooler} than the surrounding group.  
Thus, it is likely that the dark halo belongs to the group and
not to the central, giant elliptical (note that the 
group velocity dispersion profile does not decrease with radius when
the velocity and projected radial offsets are defined relative to the group
mean velocity and projected spatial centroid, as in Figure 6a).
We reach the same conclusion in Paper II by examining the X-ray images and 
spectra of the hot gas in these groups.

It is possible that contamination from
galaxies not bound to the group artifically increases the velocity
dispersion of the combined group sample, especially at large radii
from the BGG.  We can conservatively estimate the degree of
contamination by defining the 12 galaxies in the fifth (last) bin in
Figure 5 with peculiar velocities $> 1.33\sigma_{BGG}$ 
and $< -1.33\sigma_{BGG}$
as interlopers.  Because the interloper fraction should be constant
with peculiar velocity, we predict that there are a total of about
$12(3/2) = 18$ interlopers in the fifth bin.  The ratio of the areas
sampled by the fourth and fifth bins is 0.24.  Therefore, only about
four of the galaxies in the fourth bin and one or none of the galaxies
in each the inner three bins are likely to be interlopers.  Even
excluding the four galaxies with the most extreme peculiar motions
from the fourth bin only decreases the velocity dispersion in this bin
by $15\%$ to $\delta = 1.1$.  We conclude that the shape of the combined
velocity dispersion profile within $0.5$\lith\inv\ Mpc
is not significantly affected by outliers.

\subsubsection{The Mass of Group Halos}
\renewcommand{\thefootnote}{\fnsymbol{footnote}}
The virial masses of the poor X-ray groups are 
$M_{vir} \sim 0.5$-$1 \times 10^{14} h^{-1} \Mdot$.
What fraction of this mass is in the common group halo and
what fraction is associated with group galaxies?
We expect that early in a group's evolution, the
individual galaxy halos are tidally-limited by the global potential
of the group (cf. Peebles 1970; Gunn 1977).  
We must first estimate the tidal extent of the
halos of X-ray group members before we can estimate their masses.
Using Merritt's (1984) approach, we assume that the group's
potential is in the form of an analytic King 
model (King 1972)\footnote[2]{Although cosmological simulations
predict a somewhat different form for the density 
profile (Navarro \etal 1996),
we use a King profile because (1) it is a good fit to
the data (cf. Paper II) and (2) we wish to be consistent with past
analyses of rich clusters.}.\renewcommand{\thefootnote}{\arabic{footnote}}
The maximum tidal radius of a galaxy near the core of a group is
$ r_T \approx {1 \over 2} (r_c \sigma_g / \sigma_r)$,
where we adopt $r_c \sim 125$\lith\inv\ kpc
as the group's core radius (equivalent to the cluster value; Bahcall 1975)
and $\sigma_g$ as the galaxy's
internal, line-of-sight velocity dispersion.  For $L^*$ galaxies
($\sigma_g \approx 225$ \ks; Tonry \& Davis 1981) in
groups with velocity dispersions of \sigmar\ $= 200$ and 450 \ks, 
the maximum tidal radii are $\sim 75$ and 35\lith\inv\ kpc,
respectively, which are larger than the luminous radii typical 
of such galaxies.
These values of $r_T$ are less than the $> 150$\lith\inv\ kpc dark halos
of isolated disk galaxies (Zaritsky \& White 1994),
but greater than the predicted, tidally-limited
extent of galaxies in the cores of rich clusters
($\sim 15$\lith\inv\ kpc, Merritt 1984).

What is the fraction of the group mass associated with its galaxies?
We estimate the tidally-limited
mass of an $L^*$ galaxy from
the virial theorem:  $m \approx {1 \over 2} (r_T \sigma_g^2/G)$ for a
King model (1966).  For $L^*$ galaxies in a \sigmar\ $= 200$ and 
450 \ks\ group, we obtain
$m^* \sim 4\times 10^{11}$ and $2\times 10^{11} h^{-1} \ \Mdot$, respectively.
To estimate the mass of the brighter, central elliptical,
we assume that its internal
velocity dispersion is $\sigma_{BGG} \sim 300$ \ks\ (typical of the
central, giant elliptical in our X-ray groups)
and that its radius is $\sim 100$\lith\inv\ kpc (which
is between $r_T$ and $r_c$, a range 
consistent with the effective optical radii
of ``cD" galaxy envelopes (Schombert 1988), with the maximum radius to
which the X-ray emitting gas in a BGG is detected (Paper II),
and with the possibility that the BGG is less tidally-limited
than other group members, because of its position in the group center
($\S3.4$) where the tidal forces are symmetric
and relatively weak (Merritt 1984)).
We obtain $\sim 10^{12} h^{-1} \ \Mdot$ as an estimate
of the mass of the BGG's in the sample groups.
For a group like NGC 741, the mass associated with the four
$L^*$ or brighter galaxies within
the virial radius $r_h = 400$\lith\inv\ kpc is then
$\sim 10^{12} + 3(2 \times 10^{11}) h^{-1} \ \Mdot \approx 2 \times 10^{12}
h^{-1} \ \Mdot$.  

We can also estimate the fraction of
mass tied to fainter group galaxies.  Assuming that
the luminosity spectrum of group galaxies is
a Schechter function (1976) with $\gamma = 1.25$ and a lower
cut-off at $0.05L^*$ ($L^* \sim 1 \times 10^{10} \  L_{\odot}$; 
Kirshner \etal 1983), we use the empirical 
luminosity-velocity dispersion relationship ($L \propto \sigma_g^4$;
Faber \& Jackson 1976), the equation for tidal radius 
($r_T \propto \sigma_g$), and the virial theorem ($m \propto \sigma_g^2 r_T$)
to convert the luminosity function into a tidally-limited mass function
(see Merritt 1984 for a similar discussion).
We obtain
$$dn = {4 \over 3} e^{-{\left(m \over m^*\right)^{4/3}}} 
{\left(m \over m^*\right)}^{\left(1 - 4\gamma\right)/3} 
d\left({m \over m^*}\right)$$ 
for galaxies with tidally-limited masses greater than $0.1m^*$, where
we assume that the halos of
all galaxies are tidally-limited during a close
passage to the group core and that only non-luminous matter is
removed.  The mass of the $L^*$ or brighter
members is $\sim 20\%$ of the total galaxy mass.
Therefore, the total mass in galaxies in NGC 741 is
$\sim 1 \times 10^{13} h^{-1} \ \Mdot$, in contrast to the 
the virial mass of $\sim 1 \times 10^{14} h^{-1} \ \Mdot$.
The fraction of mass in the galaxies within the virial radius is then
roughly $10\%$.  In lower velocity dispersion
groups such as NGC 4325, there are fewer bright members,
but the galaxy halos are less tidally-limited and
the virial mass is smaller.  Hence,
galaxies comprise a greater fraction of the
total mass within the virial radius, $\sim 20\%$.  

The dominant source of uncertainty in the estimation of the 
fraction of group mass associated with individual galaxies is
the galaxy halo mass.  The
degree to which group galaxy halos are tidally-limited
depends on such key unknowns
as the details of galaxy halo formation, the initial orbits of the
group members, and the evolution of the group's mass density.
It is also possible that additional 
mass could be collisionally-stripped from
galaxies later in the evolution of the group (see $\S3.5$).
The exploration of these effects awaits improved simulations.

\subsubsection{Implications}
The small fraction of the group mass associated with its galaxies
will increase the timescale for galaxy-galaxy
interactions, such as mergers, close tidal encounters, 
and dynamical friction, by decreasing
the cross-sections of the galaxies (Governato \etal 1991;
Bode 1993; Athanassoula 1997).
This result argues that poor groups survive longer than 
predicted by models in which all the mass is tied
to individual galaxies and may explain why so many poor groups
are observed in lieu of single merger remnants.

\subsection{The Giant Elliptical vs. the Group Centroid}

\subsubsection{Results}
The poor groups with diffuse X-ray emission in our sample have a giant
($\simless \Mstar - 1$)
elliptical that lies within $\sim 5$-10\lith\inv\ kpc of the X-ray peak
(Paper II).
The coincidence of the BGG position and the X-ray peak
suggests that BGG's lie in the center of the group potential.
However, 
it is possible that 
X-ray emission in the core is dominated by light from the BGG,
hiding an offset between the galaxy and the center of the potential
as defined by peak of the intra-group medium.   Therefore, the optical
data are required to determine whether the BGG is at rest with respect
to the group as a whole.  In this section, we test this hypothesis
in two ways, asking (1) whether 
BGG's are more concentrated in the core than
other group members and (2) whether the radial velocity and
projected position of the BGG's are consistent with the
kinematic and projected spatial centroids of the groups.

How are the BGG's distributed in projected radius and peculiar
velocity compared with other group galaxies?
Figure 6a shows the kinematic ($y$) and projected spatial ($x$) offsets of the
BGG (filled squares) and other group
members (filled circles)
from the group centroid for the nine X-ray groups.  
Neither the projected spatial centroid ($\langle r \rangle$) nor the 
mean velocity ($\langle \upsilon \rangle$) of the group 
are weighted by galaxy luminosity, so the
velocity and projected position of the BGG do not bias the 
centroid calculations.  The velocity offset
of each galaxy from the mean velocity of the group
is divided by the group velocity dispersion \sigmar\ to compensate for
differences in the size of the group global potentials.

We define a statistic $R^2 = (x/\delta_x)^2 + (|y|/\delta_{|y|})^2$, where
$\delta_x$ and $\delta_{|y|}$ are the $rms$ deviations in $x$ and $|y|$
for the entire sample.
Thus, a galaxy that has a large peculiar motion
and/or that lies outside the projected group core will have a larger $R$ value
than a galaxy at rest in the center of the group potential.
Figure 6b shows the 
distributions of $R$ for all of the group members (solid) 
and for the subset of BGGs (shaded).
A Student's t-test gives $<3 \times 10^{-5}$ as the probability that
the means of the non-BGG and BGG distributions are consistent.
Therefore, the BGG's are significantly 
more concentrated in the core, suggesting that BGG's occupy the center
of the potential on orbits
different from those of other group members.

Are the peculiar velocities and projected positions of the BGG's
consistent with the
kinematic and projected spatial centers of the groups?
The $y$ errorbars in Figure 6a 
represent the $68\%$ confidence limits on $y$ ($\epsilon_y$) obtained
from adding the errors in the group mean velocity and the BGG velocity
in quadrature.  To estimate the $68\%$ error in $x$ 
for a BGG in a group with $N_{grp}$ members, we
use a statistical jackknife test in which samples of $N_{grp}$ galaxies
are drawn from HCG 62, the group with the most members.
For each BGG,
we adopt the $rms$ deviation in the distribution of $x$ ($\epsilon_x$)
for these samples as the $x$ error.
(For the BGG in HCG 62, we use the smallest $x$ error calculated
among the other groups).
The heavy line in Figure 6b shows the distribution of
$R$ obtained by assuming that all of the BGG's lie in the centers of 
their groups.  For each BGG,
we model the distributions of errors about the center in $|y|$
and in $x$ as
Gaussian with $rms$ deviations equal to $\epsilon_y / \delta_{|y|}$
and $\epsilon_x / \delta_x$, respectively.
The heavy line is the distribution in $R$ that 
results from 1000 random draws from the
model distributions of $x$ and $|y|$
for each BGG.  A t-test fails to differentiate
between the means of the model $R$ distribution and that of the BGGs (shaded)
at better than the $95\%$ confidence level.
We conclude that, to within the errors, the BGG's occupy
the center of the group potential.

\subsubsection{Timescales for BGG Evolution}
The presence of a giant elliptical at rest 
in the center of each X-ray group is consistent with the picture in which
BGG's form from galaxy mergers early in the group's evolution
(Merritt 1985; Bode 1994).  BGG formation may occur
during the initial collapse of the group,
before galaxy halos are tidally-truncated and merger rates decline
(Merritt 1985).
The absolute magnitudes of the BGG's place them in
the class of bright ellipticals whose morphologies and kinematics
are consistent with merger evolution ($M_B \simless -20 +5$log$_{10}$ \lith;
Kormendy \& Bender 1996; Merritt \& Tremblay 1996).
Such a galaxy may be forming in HCG 90.
The total luminosity of the several
merging galaxies in HCG 90's core is comparable to the
absolute magnitudes of the BGG's in the other X-ray groups. 

Do BGG's continue to experience mergers after their formation?
The BGG's in our sample show no signatures of later mergers with 
massive galaxies; their spectra
do not indicate recent star formation, their morphologies are not obviously
disturbed, and their position in the center of the
X-ray emission argues against recent disruptions of the
surrounding gas by mergers.
For example, one BGG (NGC 5846) is classified as an intermediate age elliptical 
(the last star formation 
event occurred more than $\sim 6$-7\lith\inv\ Gyr ago; Trager 1997).
However, NGC 2300, the central, giant elliptical
in the first poor X-ray group discovered
(Mulchaey \etal 1993), has shells (Schweizer \& Seitzer 1992; 
Forbes \& Thomson 1992)
and a spectrum that
includes a young stellar population
($\simless 2$\lith\inv\ Gyr old;
Trager 1997).  Thus, there is some indirect evidence that at least one BGG
has experienced a late merger.

We can estimate whether the central elliptical is likely to grow
via subsequent mergers with other group members.
One possible mechanism is the accumulation of slow-moving, massive galaxies
in the group core by dynamical friction.
To determine the effects of dynamical friction, we roughly
estimate the largest radius from which an $L^*$ galaxy could fall into the center of a group
in a Hubble time.  We use the dynamical friction timescale (Binney \& Tremaine
1987; their eq. 7-26) and the Coulomb logarithm definition (7-13b) 
for an $L^*$ galaxy on an initially circular orbit to solve for the maximum radius.
We use the tidally-limited masses
of an $L^*$ galaxy in a \sigmar\ $= 450$ \ks\ and 200 \ks\ 
group from $\S{3.3}$.
For a galaxy moving with circular velocity $\upsilon_c = \sqrt 2$\sigmar, 
the maximum radius ranges from $\sim 250$\lith\inv\ kpc for the high velocity dispersion group to
400\lith\inv\ kpc for the low \sigmar\ group.  
This timescale estimate is consistent
with simulations (Merritt 1984), which predict that dynamical
friction can cause 1-$2L^*$ worth
of merger candidates 
to accumulate in the core of a \sigmar\ $= 500$ \ks\ group 
within a Hubble time.
If these candidates do merge with the BGG, some 
enhancement of the BGG's luminosity is possible, and
the luminosity segregation resulting from dynamical friction
might then be hidden by the mergers.
In contrast, the same equations and simulations predict that
an $L^*$ galaxy in a rich cluster
($\sigma_r \sim 1000$ \ks) requires 
an improbable $\sim 40$\lith\inv\ Gyr to fall to the center from
an initial radius of 250\lith\inv\ kpc.

\subsubsection{Implications}
The existence of a central, giant elliptical in the X-ray groups
has implications for the formation of ``cD" cluster 
galaxies (Matthews, Morgan, \& Schmidt 1965).  The method by which ``cD"s
evolve is unknown, but there is indirect evidence that these galaxies
form outside the cores of rich clusters.  For example,
``cD"s are not always at rest with
respect to the cluster potential, but lie instead 
in local overdensities (Beers \& Geller 1983; Beers \etal 1995).  
In addition, the presence of a ``cD" does not depend
on the global kinematic properties of the cluster ({\it i.e.,}
clusters that contain a ``cD" have similar velocity dispersions to those that
do not, Zabludoff \etal 1990).  
Our observations suggest that ``cD"s
form from galaxy-galaxy mergers in poor groups.
Because poor group
velocity dispersions are comparable to the internal
velocity dispersions of their galaxies, the probability that a
galaxy-galaxy collision will lead to a merger is higher in groups
than in clusters (see $\S3.5$).  
Poor groups with central, massive ellipticals may later merge to form clusters
or fall into existing clusters (Merritt 1985; Beers \etal 1995).  
In this picture,
the BGG would be displaced for a time from the center of the cluster potential.
We note that the NGC 2563 X-ray group is one of the groups in Cancer
(Bothun \etal 1983), an association of groups
likely to evolve into a rich cluster after its virialization.
In future work, we will use deep optical surface photometry
to determine whether the
BGG's in the X-ray groups have the extended halos of ``cD"
galaxies in rich clusters.

\subsection{Group Early-Type Fractions}

\subsubsection{Results}

The effects of environment on galaxy evolution are poorly understood.
The study of poor groups provides a critical link between
the evolution of isolated galaxies in the field and of galaxies
subject to the hot, dense environments of clusters.
Despite the usefulness of group galaxies 
as a control sample, their properties, especially at faint magnitudes,
are not well-known.
One possible test of environmental influences
is to compare the morphologies of group members with
those of field and cluster galaxies.  Previous work suggests
that the fraction of early type (E and S0) galaxies in X-ray detected 
groups varies widely and that some X-ray groups 
have no late types among their brightest members
(Ebeling \etal 1994; Pildis \etal 1995; 
Henry \etal 1995; Mulchaey \etal 1996b).  
However, these studies typically include only the four
or five brightest galaxies, which biases samples toward ellipticals,
and target only the central $\simless 0.3$\lith\inv\ Mpc,
where early types concentrate ({\it e.g.}, Figure 3).  Therefore,
to compare the morphologies of group 
and cluster members, we must sample both environments to similar
physical radii and absolute magnitude limits.

We plot the distribution of the early type (E, S0) fractions $f$ of 
the 12 sample groups in Figure 7.
In the X-ray groups, $f$ ranges widely from $\sim 55\%$
(HCG 62, NGC 741, and NGC 533) 
to $\sim 25\%$ ({\it e.g.}, NGC 2563).  The lower value of $f$
for the X-ray groups
is characteristic of the field ($\sim 30\%$; Oemler 1992).
We find no early types among the 6-8 galaxies in each of
the three non-X-ray groups (shaded).

We can test whether the groups with the
highest early type fractions, such as NGC 533 (14 of 25)
and NGC 741 (10 of 18), are statistically different from the field.
The likelihood of at least $k$ successes in $n$ Bernoulli trials with success
probability $f_F$
is $\sum_j {n! \over {j!(n-j)!}} {f_F}^j (1 - f_F)^{n-j}$ 
for all $j$ such that $n \geq j \geq k$.
The fraction of field galaxies that are
E's and S0's ($f_F$) is $\sim 30\%$ (Oemler 1992).
If we assume that the
E and S0 fraction in NGC 533 is actually consistent with $f_F$,
the probability of finding 14 or 
more early types in a sample of 25 group members
is very small, $6 \times 10^{-3}$.  The probability that
the early type fraction in NGC 741 is consistent with the field is
also small, $2 \times 10^{-2}$.  
When these probabilities are considered together,  
the hypothesis that
the early type fractions of both groups are consistent with the field
is rejected at the 
$\sim 4 \sigma$ level.  NGC 533, NGC 741, and HCG 62 are sampled
to physical radii and absolute magnitude limits similar to
NGC 2563, an X-ray group in which $f$ is consistent with the field,
and to NGC 644, a non-X-ray group in which $f$ is much lower than the 
field ($0\%$).  We conclude that
the spread in the distribution of $f$ in Figure 7 is statistically
significant.

How do the early type fractions of $\sim 55\%$ in 
NGC 533, NGC 741, and HCG 62 (16 of 30)
compare with values typical of rich clusters?
These three groups are sampled to
physical radii of $\sim 0.6$-0.8\lith\inv\ Mpc, and their members are typed to
absolute magnitude
limits of $M_B \sim -16$ to $-17 + 5$log$_{10}$ \lith.  
Within this range of radii, 
the early type fractions of rich clusters are $\sim 0.55$-0.65
(Whitmore \etal 1993).  Because  
the absolute magnitude of the Whitmore \etal sample 
($M_B \sim -18 + 5$log$_{10}$ \lith) is
brighter than ours, the cluster early type fractions are relatively biased
toward ellipticals.  Cutting the group data at the brighter
$M_B$ limit worsens the statistics, but increases $f$.
For example, six of the seven galaxies in NGC 533 brighter than
the Whitmore \etal limit are E's or S0's.
Likewise, five of the remaining six members of NGC 741 are early types.
Therefore, it is conservative to conclude that
the early type fractions in some poor groups are comparable to
those in rich clusters.

\subsubsection{The Early Type Fraction-Velocity Dispersion Relation}
The inset in Figure 7 shows the correlation between early type fraction and
group velocity dispersion for the 12 sample groups.
The solid line is an unweighted fit to the data, the dashed line
is a fit weighted by the velocity dispersion 
errors.  In both
cases, the correlation between morphology and
velocity dispersion is significant at the $>0.999$ level.
The relation is surprisingly robust, given that the groups
are sampled over a range of physical radii ($\sim 0.2$-1.0\lith\inv\ Mpc)
and absolute magnitude limits ($M_B \sim -14$ to $-17 + 5$log$_{10}$ \lith).  
It is possible that
the relation is artificially strengthened by the three non-X-ray 
groups, which may not be bound.  We note, however, that the
Local Group would similarly anchor the tail of the correlation.
The form of the relation cannot be the same for rich clusters ---
our fit to the group data predicts that the early type fraction
in a $\sigma_r \sim 1000$ \ks\ cluster is an unphysical $f = 124\%$.
Therefore, the relation must turn up between the poor group and rich
cluster regimes.  The group $f - \sigma_r$ relation implies either
that galaxy morphology is set by the local potential size at the
time of galaxy formation (Hickson, Huchra, \& Kindl 1988) and/or that 
\sigmar\ and $f$ increase as a group evolves (Diaferio \etal 1995).

\subsubsection{Galaxy-Galaxy Collision Timescales}
If environment does alter the morphologies of group galaxies
after the formation of the group, which environmental influences
are most important?  Proposed mechanisms to disturb the 
distribution of stars
in cluster galaxies include the tidal shocking of stellar disks by
the global potential (B. Moore 1997, priv. comm.), 
galaxy-galaxy harassment (Moore \etal 1996), 
collisional stripping ({\it e.g.}, Richstone 1975), and mergers ({\it e.g.},
Barnes 1989; Weil \& Hernquist 1996).
Strong tidal shocks could
strip the stellar disks from Sc or Sd galaxies passing through the strong
gravitational field of the group
core, producing remnants consistent
with the mass profiles of 
dwarf elliptical galaxies (B. Moore 1997, priv. comm.); however, 
only four of the 72 galaxies in our early type sample
are as faint as dE/dSph's ({\it i.e.,} 
$M_B \simgreat -16 + 5$log$_{10}$ \lith), 
so we do not consider this mechanism in interpreting the group data.
Galaxy harassment is not likely to operate effectively in groups, where
the number density of bright galaxies and the global velocity
dispersion are much smaller than in clusters (B. Moore 1997, priv. comm.).
In the following discussion, we
calculate the likelihood of
galaxy-galaxy encounters (collisions, mergers) by adapting simple models
applied in the past to rich clusters (cf. Richstone \& Malumuth 1983;
Merritt 1984).

Galaxy-galaxy interactions are known to significantly affect galaxy 
morphology ({\it e.g.}, Schweizer 1986, Hibbard \etal 1994).
Observations of poor compact groups suggest that
close tidal interactions and accretion events
play a role in the evolution of some group galaxies.
For example, the isophotes of compact group ellipticals 
are typically more irregular than those of ellipticals in
cluster environments (Zepf \& Whitmore 1993), and
the fraction of group members with disturbed
morphologies and peculiar kinematics is larger than for
field galaxies (Rubin \etal 1991; de Oliveira \& Hickson 1994).
If poor groups are longer-lived than previously supposed,
and merged galaxies have had more time to relax, the
fraction of merger remnants in groups may be even higher than indicated
by short-lived merger signatures.

What is the timescale for galaxy-galaxy encounters in
our poor groups?  We now examine the timescales in four
cases:  (a) an $L^*$ galaxy in a \sigmar\ = 450 \ks\ group,
(b) a galaxy of typical luminosity $L_{typ}$ in a \sigmar\ = 450 \ks\ group,
(c) an $L^*$ galaxy in a \sigmar\ = 200 \ks\ group, and 
(d) an $L_{typ}$ galaxy in a \sigmar\ = 200 \ks\ group.

{\bf Definition of Collision Timescale.}
Following Richstone \& Malumuth (1983), we define the collision time
as $T_{col} = [{\overline n} \pi {r_T}^{2} \sqrt{2} \sigma_r]^{-1}$,
where ${\overline n}$ is the mean number density of galaxies
within the half-mass radius, ${r_T}$ is the tidal radius of the galaxy,
and $\sqrt{2} \sigma_r$ is the typical encounter velocity in
an isotropic system.  We assume that the group mass distribution
inside the virial radius
is described by a King model (1966) with a core radius of
$r_c \sim 125$\lith\inv\ kpc and with a shape defined by the ratio
of the group tidal radius to $r_c$,
$R_T/r_c = 20.2$ (see Bahcall 1977 and Dressler 1978 for discussions
of the shapes and sizes of clusters).
The half-mass radius of this model is $2.83 r_c = 354$\lith\inv\ kpc,
comparable to the typical
group virial radius.  Therefore, we calculate ${\overline n}$ using the
mass within the virial radius.  The  
mean mass of a group member for the tidally-limited mass function
defined in $\S3.3$
is ${\overline m} = 0.35m^*$, where $m^*$ is the mass of an $L^*$ galaxy.

{\bf (a) $L^*$ Galaxy in a \sigmar\ = 450 \ks\ Group.}
To calculate $T_{col}$ for an $L^*$ galaxy in a \sigmar\ = 450 \ks\ group,
we use the NGC 533 group as an example.  In $\S3.3$, 
we found $r_T \sim 35$\lith\inv\ kpc for an $L^*$
galaxy in this group.  What is ${\overline n}$?
The total mass of NGC 533 within
$r_h = 400$\lith\inv\ kpc is $M_{vir} \sim 1 \times 10^{14} h^{-1} \ \Mdot$.
Thus, if we assume that NGC 533 is spherical and 
that $f_{halo} \sim 90\%$ is the fraction of the
total mass not tied to the galaxies, then the
average galaxy number density within $r_h$ is
${\overline n} = (3 M_{vir} / 
4 \pi {r_h}^3) (1 - f_{halo}) / {\overline m} \sim 529 h^3$ Mpc$^{-3}$.
This calculation shows that
we would overestimate ${\overline n}$ and thus
underestimate the collision timescale by a factor of 10 by
assuming that all of the group mass is associated with
the galaxies.  Note also that
$f_{halo}$ is assumed to be constant over time, as would be the case
if the tidal limitation of the galaxies occurred early in the group's
evolution.  If $f_{halo}$ increases in time, then the collision timescales
below are somewhat overestimated.  

With these values of ${\overline n}$ and $r_T$,
we obtain $T_{col} \sim 0.07\times$ a Hubble time ---
an $L^*$ galaxy experiences a collision in every $\sim$
four crossing times.  In comparison, a rich cluster like Coma has
at least $\sim 5 \times$ the galaxy number density and more than
twice the velocity dispersion of NGC 533.  If cluster galaxies
have halos truncated to about half the extent of those in NGC 533 (Merritt 1984),
then the collision time for an $L^*$ galaxy in a rich cluster is
$\sim 0.04$ of a Hubble time.
Therefore, over the same period of time,
an $L^*$ galaxy in a rich cluster is nearly
twice as likely to experience a collision as an $L^*$ galaxy
in NGC 533, the poor group with the shortest 
estimated $T_{col}$ in the sample.

{\bf (b) $L_{typ}$ Galaxy in a \sigmar\ = 450 \ks\ Group.}
To understand more about how the collision history of a galaxy depends
on its luminosity and on the velocity dispersion of the group,
we estimate the collision time for a typically bright
galaxy in NGC 533.  The value of ${\overline n}$ is the same as in case (a),
but what is $r_T$?  The luminosity of a typically bright galaxy
is $L_{typ} \approx 0.15L^*$ for the Schechter luminosity function 
assumed in $\S3.3$ (Richstone \& Malumuth 1983).
$L_{typ}$ is two magnitudes fainter than $L^*$ and is
also the approximate limit at which the completeness of
our group samples is at least $\sim 60\%$ ($\S3.1$).
We use 
the Faber-Jackson (1976) relation and the internal velocity dispersion
of an $L^*$ galaxy ($\approx 225$ \ks; Tonry \& Davis 1981)
to estimate $\sigma_g$ for an $L_{typ}$ galaxy
($\sim 140$ \ks).  The tidal radius of
an $L_{typ}$ galaxy is then $\sim 20$\lith\inv\ kpc,
and the collision time is about three times longer than for
an $L^*$ galaxy in NGC 533.  

{\bf (c) $L^*$ Galaxy in a \sigmar\ = 200 \ks\ Group.}
An $L^*$ galaxy in a lower velocity dispersion group will collide
less frequently than will its counterpart in NGC 533.
The tidal radius of this galaxy is $\sim 75$\lith\inv\ kpc ($\S3.3$).
What is ${\overline n}$?
For a \sigmar\ = 200 \ks\ group, we estimate
the virial mass within 400\lith\inv\ kpc by 
assuming that the virial radii are the same and scaling NGC 533's $M_{vir}$
by the square of the
ratio of the velocity dispersions of the two groups.
Because (1) the virial mass is smaller, (2) the 
average galaxy mass is larger, and (3) 
$f_{halo} \sim 80\%$ is smaller
in the \sigmar\ = 200 \ks\ group,
${\overline n}$ is a factor of $\sim$five times smaller than in NGC 533.
As a consequence, the collision timescale
for an $L^*$ galaxy in a \sigmar\ = 200 \ks\ group 
is $\sim 3\times$ longer
than for an $L^*$ galaxy in NGC 533.

{\bf (d) $L_{typ}$ Galaxy in a \sigmar\ = 200 \ks\ Group.}
The tidal radius of a typically bright galaxy in a
\sigmar\ = 200 \ks\ group is $\sim 45$\lith\inv\ kpc, where we
assume that the galaxy's velocity dispersion is the same as in case (b).
The value of ${\overline n}$ is the same as in case (c).
Therefore, $T_{col}$ is $\sim 8\times$ longer in this case than for
an $L^*$ galaxy in NGC 533.

{\bf Summary.}  Collision timescales are shortest
for bright cluster members,
whose cross-sections are large because of the high relative velocities
and galaxy number density.  $T_{col}$ is increasingly longer for
an $L^*$ galaxy in a \sigmar\ = 450 \ks\ group, 
an $L^*$ galaxy in a \sigmar\ = 200 \ks\ group, 
an $L_{typ}$ galaxy in a \sigmar\ = 450 \ks\ group, 
and an $L_{typ}$ galaxy in the lower \sigmar\
group.  
However, as we will see, the merger timescales do
not follow this hierarchy, because they depend
not only on the collision timescale, but also on the
fraction of collisions that lead to mergers.

\subsubsection{Galaxy-Galaxy Merger Timescales}
To roughly calculate the merger timescale, we
assume that galaxies in collision merge if the ratio
of the collision speed to their internal velocity dispersion ($\sigma_g$)
is $\leq 3$ (Tremaine 1980; Richstone \& Malumuth 1983).
In other words, we determine
the fraction of collisions $\zeta$ that are
slower than $3\sigma_g$.  As discussed in Richstone \& Malumuth (1983),
the fraction of relative velocities in one direction varies
as $e^{-\upsilon/4{\sigma_r}^2}$.  The merger fraction
is then $\zeta = \int_{0}^{3\sigma_g} 
e^{-\upsilon^2/4{\sigma_r}^2}\upsilon^2 d\upsilon / \int_0^{\infty}
e^{-\upsilon^2/4{\sigma_r}^2}\upsilon^2 d\upsilon$.
The merger timescale is thus $T_{merg} = T_{col} \zeta^{-1}$.
For an $L^*$ galaxy in NGC 533, 
$T_{merg} \sim 4T_{col} \sim 0.3t_H$.
In contrast, the merger timescale
in a rich cluster is $\sim 100\times$ its 
collision time (Richstone \& Malumuth 1983), or
about four Hubble times.  Thus, $T_{merg}$ for an $L^*$ galaxy
is $\sim 13\times$ shorter in NGC 533 than in rich clusters.

For an $L_{typ}$ galaxy in NGC 533, the ratio of the
velocity dispersion of the group to that of the galaxy
is larger.  Therefore, the fraction of collisions that lead
to mergers is smaller, and $T_{merg}$ is $11\times$ longer,
than for an $L^*$ galaxy.
The efficiency of mergers is
higher in a \sigmar\ = 200 \ks\ group, because
the internal velocity dispersion of an $L^*$ and an $L_{typ}$ galaxy
is comparable to \sigmar.  This effect is somewhat offset by
fewer collisions in
the lower velocity dispersion group.  Therefore, the merger
timescale for an $L^*$ and an $L_{typ}$ galaxy is
$0.7\times$ and $4\times$ that for
an $L^*$ galaxy in NGC 533, respectively.

In summary, the merger timescales are increasingly longer for
an $L^*$ galaxy in a \sigmar\ = 200 \ks\ group, 
an $L^*$ galaxy in a \sigmar\ = 450 \ks\ group, 
an $L_{typ}$ galaxy in the lower \sigmar\
group, an $L_{typ}$ galaxy in the higher \sigmar\ group,
and an $L^*$ galaxy in a rich cluster.

\subsubsection{Collisional Stripping Timescale}
Even if two colliding galaxies do not merge, the encounter
may disrupt their morphologies.
The timescale over which mass is removed from a colliding galaxy
is shorter in poor groups
than in clusters, because the slower relative velocities of the
colliding galaxies makes stripping more effective.
Numerical experiments
(Richstone 1975; Dekel, Lecar, \& Shaham 1980; Richstone \& Malumuth 1983) 
suggest that only 1-10\% of a galaxy's mass is lost during a collision at
velocities typical of galaxies in rich cluster.  If we employ
the upper bound on the fraction of mass lost,
then $T_{strip} \sim 10 T_{col}$ is a very rough estimate of the
collisional stripping timescale in our highest velocity dispersion groups.
Therefore, a collision that is effective at stripping mass from an
$L^*$ galaxy in NGC 533 might occur within a Hubble time.
More sophisticated simulations are required before this estimate
can be improved.

\subsubsection{Implications}
What do the merger timescales
imply for galaxy evolution in poor groups?
If \sigmar\ = 450 \ks\ groups like NGC 533 form
from the mergers of lower \sigmar\ groups, galaxy-galaxy
mergers occur less frequently as the group evolves.
In this case, we would expect to observe more on-going
galaxy mergers in lower \sigmar\ groups and 
more merger remnants in higher \sigmar\ groups.
In a sample of compact groups with velocity dispersions of order 200 \ks,
a high fraction of galaxies
are interacting ($\sim 43\%$; de Oliveira \& Hickson 1994).
The early type fractions in our $\sigma_r \simgreat 400$ \ks\ groups
are as high as in rich clusters.
If some early type galaxies are evolved merger remnants, 
this highly-simplified model is qualitatively
consistent with the observed relationship between
early type fraction and velocity dispersion, {\it i.e.,} the
galaxy populations of
higher velocity dispersion groups are more evolved on average.
At some point in the group's evolution, perhaps at a
velocity dispersion near 400 \ks, any morphological
evolution resulting from galaxy mergers ceases,
and the fraction of merger remnants in poor groups and rich clusters
is comparable.  The implicit upturn in our $f - \sigma_r$ relation
suggests such a saturation point.

Alternatively, the similarity of some group and cluster
early type fractions, and the steepening of 
the $f - \sigma_r$ relationship at high \sigmar,
might arise from conditions at the time of galaxy formation.  For example,
it is possible that poor groups such as NGC 533 and rich clusters like
Coma begin as similar mass density perturbations
with correspondingly similar
galaxy populations.  In this simple picture, NGC 533 does not develop a
cluster-size potential, because its field lacks the surrounding
groups that Coma later accretes.  

In summary, the cluster-like fraction of early type galaxies
in NGC 533, NGC 741, and HCG 62 suggests that the cluster environment is
not required to produce copious quantities of E and S0 galaxies.
The simplest explanation is either that fluctuations in
the initial conditions permitted early types to form
in these comparatively low velocity dispersion, low galaxy density
environments or that the galaxies' subsequent evolution
was the product of a mechanism,
such as galaxy-galaxy interactions,
common to both groups in the field and groups that become subclusters.
Although additional environmental 
mechanisms may affect the evolution of cluster galaxies, such
cluster-specific processes are not required to explain the current data.
A cluster that evolves hierarchically from subclusters with the properties
of NGC 533, NGC 741, and HCG 62 will have, at least initially, a similar
galaxy population.  As we demonstrate below,
another test of the importance of cluster
environment on galaxy evolution at the current epoch
is to compare the
recent star formation histories of galaxies in poor groups and
in subclusters.

\vfill\eject
\subsection{Early-Type Group Galaxies with Young Stellar Populations}

\subsubsection{Results}
The star formation histories of galaxies in poor groups
provides additional insight into the 
environmental factors that may influence the evolution of galaxies.
One approach 
is to examine the spectra of the early types for evidence of
on-going star formation or of a young stellar population.
We can then compare the fraction of 
E and S0 group members that have recently formed stars with
a sample from rich clusters with complex structure.
If early types in poor field groups
have different recent star formation histories than those
in infalling subclusters, which have only just been exposed
to the cluster environment for the first time,
then cluster-based galaxy evolutionary mechanisms
are clearly potent at the current epoch.
On the other hand, if the two samples are similar,
then the mechanisms that operate exclusively
in clusters or that are much more effective in clusters
than in poor groups are not required to 
enhance and/or quench
star formation in these galaxies.  In the latter case,
recent star formation may be the product of galaxy-galaxy
encounters, which are present in both groups and subclusters,
or of evolution that is independent of the galaxies' present environment.

To determine which of the E and S0 group members have 
spectra that indicate on-going or recent star formation,
and to compare this population with that in rich, complex clusters,
we first automate the calculation of the
[O II]$\ \lambda 3727$
equivalent width in the manner of Zabludoff \etal (1996).
The equivalent width uncertainties, which are typically less than 1 \AA, 
are calculated
using counting statistics (the detector is a photon counter with
approximately zero read noise), the local noise in the
continuum, and standard propagation of errors.  
Figure 8a shows the unfluxed spectra of
the central regions of four early type group
members with significant [OII]
emission ($> 5$\AA, excessive for a normal Sa type galaxy
(Kennicutt 1992ab; Caldwell \& Rose 1997)).  
This criterion is the same as that used
by Caldwell \& Rose (1997) to identify star
formation in the spectra of early type galaxies in rich clusters
with substructure.  
However, the ratios of
[OIII]$\ \lambda 5007$ to H$\beta \ \lambda 4861$ and
[OII]$\ \lambda3727$ to [OIII]$\ \lambda 5007$ suggest that
H$90\_{017}$ is
an AGN. Two other spectra in Figure 8a, H$42\_{023}$ and H$62\_{008}$,
are either star forming or weaker AGN.  To be conservative, we include only
N$741\_{020}$ as a star forming galaxy in the subsequent analysis.

Are there early-type group members that
have experienced star formation in the last few Gyr
and are now quiescent?
The ratio of the strength of the 
Ca II H $\lambda 3968 \ +$ Balmer H$\epsilon \ \lambda 3970$ line to the
Ca II K $\lambda 3934$ line is a strong indicator of
recent star formation (cf. Leonardi \& Rose 1996), especially when
there is some filling of the other Balmer absorption lines.  
We visually identify
the spectra of group galaxies with early type morphologies that
have H+$H\epsilon$ to K ratios $> 1$ and signal-to-noise
ratios $S/N > 6$.  These seven spectra are shown in Figure 8b.

To further test 
whether the galaxies in Figure 8ab have a young stellar component,
we calculate the 4000 \AA\ break $D_{4000}$ for the sample spectra
as in Zabludoff \etal (1996).  The break strength is
a measure of the galaxy's color, with lower $D_{4000}$ indicating a
bluer stellar population.
The eight star forming and post-star forming galaxies in Figure 8ab
lie in the blue
tail of the $D_{4000}$ distribution (Figure 9, shaded), implying that
they have a younger stellar population
than is typical for the other early types in the sample (white).
The unshaded galaxy blueward of the peak of the shaded subsample
is the probable AGN discussed above, H$90\_{017}$.
The case for recent star formation in the shaded subsample is
bolstered by stellar population synthesis models 
(Bruzual \& Charlot 1995),
which indicate that for a double burst model of
star formation in which the second burst lasts a Gyr or less,
a galaxy with $D_{4000} \simless 2$ has experienced the second
burst within the last $\sim 2$ Gyr,
regardless of the assumed initial mass function, burst
strength, or progenitor type.

\subsubsection{Comparison with Rich Clusters}
The star forming and post-star forming galaxies whose spectra 
are in Figure 8ab are $12\%$ of the
64 early type group members for which we have spectra.
How does this fraction compare with that in clusters with 
infalling groups?  Caldwell and Rose (1997) examined galaxy spectra 
in clusters with significant substructure 
and found that $\sim 15\%$ of galaxies with early type morphologies
have signatures of recent star formation.
The resolution of our spectra
is $\sim 3\times$ poorer than for that cluster sample,
preventing us from
precisely duplicating the young stellar population
index used by Caldwell and Rose ({\it e.g.},
their continuum definitions correspond
to one pixel or less in our spectra).
However, like all of our post-star forming early types,
most of the galaxies in Caldwell and Rose's post-starburst sample
have sufficiently strong Balmer $H\epsilon$ so that
H+$H\epsilon$ EW $>$ K EW (note that Caldwell and Rose include
a few possible AGN and transitional S0/a type galaxies of
the kind that we exclude from the group sample).
Therefore, the fractions of early types with
``abnormal" Balmer absorption and/or [OII] emission
in poor groups and in subclusters are roughly consistent.

There are two potential problems in comparing the poor group and
cluster samples.  First, although the spectroscopy of both
samples was conducted with fibers of similar size,
the physical radius to which the group and cluster galaxies
are sampled by the fibers differs, because
the Caldwell and Rose systems are generally more distant than ours.
This ``aperture bias" might alter the apparent contributions of the
young and old stellar populations in our
our spectra relative to the Caldwell and Rose data.
For example, our fibers will tend to miss [OII] emission or Balmer absorption
in the outer parts of some E and S0's and to oversample nuclear light
in comparison with the Caldwell and Rose spectra.
Second, our group sample includes early type galaxies that are
several magnitudes fainter than those
in the cluster sample.  
Although a Kolmogorov-Smirnov test fails to distinguish (at better than the
$95\%$ level) 
between the distributions of estimated absolute magnitude
$M_B$ for the early types with and without
young stellar populations in our groups, an
increase in the fraction of recently star forming galaxies at fainter
absolute magnitudes would complicate the comparison of the group and cluster
samples.

One way to roughly address these two issues is to compare
the Caldwell and Rose data in Coma with our data in NGC 533, because these
systems are at similar distances.  We cut the NGC 533 sample at 
the absolute magnitude limits of the Coma sample,
$M_B \simless -16.6 + 5$log$_{10}$ \lith.
Three of the 16 galaxies that 
that we classify as early types have young stellar populations,
but no on-going star formation.
This fraction of $23$\% is not significantly different from
the fraction of early types with post-starburst spectra
in the NGC 4839 subcluster ($11/38 = 29\%$; Caldwell \& Rose 1997),
whose non-coincident galaxy and X-ray gas distributions suggest that the group
is falling into the Coma cluster for the 
first time (Caldwell \etal 1993; 
White \etal 1993a; Zabludoff \& Zaritsky 1995).
Although the statistics of the samples are poor,
it is suggestive that the early type galaxies in NGC 533
have a recent star formation history similar to those
in the NGC 4839 subcluster, which has the highest
fraction of post-starburst early types
of any subclump in a nearby cluster.

The spectra in Figure 8b indicate that some of the early type galaxies
in poor groups experienced recent episodes of star formation and are
now quiescent.
What processes might induce and/or quench star formation in a
group galaxy?  
Mechanisms proposed to deplete gas in rich cluster galaxies include
ram pressure stripping (Gunn \& Gott 1972), transport processes like 
viscous stripping and thermal conduction (Nulsen 1982; Cayatte \etal 1994), 
and the expulsion of gas via supernovae-driven winds (cf.
Larson \& Dinerstein 1975).
The tidal limitation of a galaxy's halo by the group potential
($\S3.3$) might also remove gas from a reservoir outside the 
optical radius.  By removing gas from galaxies, these processes
are likely to reduce subsequent star formation.  Although
there are no models at present, it is also conceivable that these
mechanisms compress or shock inter-stellar gas, producing an increase
in star formation.  Because the metallicity
of the intra-group gas is poorly known, it is difficult to place strong
constraints on the contribution of supernovae ejecta to the
intra-group medium.  Therefore,
we consider only the effects of ram pressure stripping and
gas transport processes below.

\subsubsection{The Effects of Ram Pressure}
It is possible to roughly estimate the ram pressure of the intra-group
medium on the inter-stellar medium of group members.  Specifically, we
estimate the radius $r_s$ at which the disk of an $L^*$ galaxy 
is subject to stripping.
The condition for stripping for a galaxy that moves with uniform
velocity through a uniform intra-group medium 
is $\rho_{IGM} {\upsilon_{\perp}}^2 > 2 \pi G \sigma_{tot} \sigma_{gas}$,
where $\rho_{IGM}$ is the density of the intra-group medium,
$\upsilon_{\perp}$ is the component of galaxy velocity relative
to the intra-group medium 
that is perpendicular to the disk, $\sigma_{tot}$ is the 
surface density of the disk at radius $r_s$, and $\sigma_{gas}$ is the surface
density of gas in the disk at $r_s$ (Gunn \& Gott 1972).
This simple condition is consistent with the predictions of
n-body/hydrodynamic simulations (Kundi\'c \etal 1997).
Because NGC 533 has the highest $\rho_{IGM} {\upsilon_{\perp}}^2$ 
of the 12 groups, we estimate the
ram pressure experienced by one of its late type members to obtain a
conservative upper limit for the sample.

First we calculate the ram pressure term for a galaxy in the core of NGC 533.
We adopt $\upsilon_{\perp} \approx \sqrt 2 \sigma_r =$ 650 \ks.
We integrate a King profile (King 1966) of NGC 533's intra-group medium
that excludes the central elliptical (Paper II)
to obtain an estimate of the central gas density,
$\rho_{0,IGM} \sim 2 \times 10^{-3} h$ cm$^{-3}$.
The average gas density 
within 300\lith\inv\ kpc is then 
$\overline {\rho_{IGM}} \sim 9 \times 10^{-4} h$ cm$^{-3}$.
Therefore, the ram pressure on a galaxy in the core
of NGC 533 is
$\overline {\rho_{IGM}} {\upsilon_{\perp}}^2 \sim 4 \times 10^{12} h$ 
cm$^{-1}$ s$^{-2}$.

To determine the radius in a disk galaxy at which ram pressure
stripping is effective, we calculate where
the restoring force pressure
$2 \pi G \sigma_{tot} \sigma_{gas}$ exceeds the ram pressure.
We use the averaged
total mass and HI surface density profiles of six nearby, 
$L^*$ or brighter galaxies
from Giovanelli \etal (1981)
to obtain estimates of $\sigma_{tot}$ and $\sigma_{gas}$ at different
radii (note that the units in their Figure 7
are mislabelled $\Mdot$ kpc$^{-2}$, instead
of $\Mdot$ pc$^{-2}$).  The exponential scale length of the
total mass in the composite disk is $\sim 4$ kpc.  The
ram pressure condition is satisfied at $r_s \sim 15$ kpc, 
or about four scale lengths.  This radius is comparable to the furthest extent
of optical light in disk galaxies ($\sim 3$-5 scale lengths;
van der Kruit \& Searle 1981).  As a result, ram pressure
will probably not strip gas inside the optical disk of a
$L^*$ group member, but may deplete gas if a reservoir exists at
larger radii, limiting subsequent star formation by preventing 
replenishment by infalling gas.  It is less likely that
ram pressure will affect the morphology of group galaxies, {\it e.g.},
transforming the bulge-to-disk ratio of a late
type spiral into that of an S0.

We note that the estimated central gas density in NGC 533
is comparable to that in Coma, where
$\rho_{0,ICM} \sim 5 \times 10^{-3} h^{-1/2}$ cm$^{-3}$ (Hughes 1989).
If we assume conservatively that
the {\it average} gas density is the same in groups and clusters,
and thus
that differences in ram pressure depend only on global velocity dispersion,
then galaxies in a \sigmar\ $\sim 1000$ \ks\ rich cluster
experience at least $\sim 5$-25$\times$ more ram pressure on average than do
the members of our groups.

There are several limitations of this estimate of
the ram pressure in poor groups.  First, the calculation does not include
the replenishment by supernovae of gas removed by ram pressure
from the outer disk and halo.  Second, we
assume that gas is uniformly distributed in the disk ---
any clumping makes the gas 
harder to strip away because the stripping force is reduced in proportion
to the area of the gas clump, while 
the gravitational force is unaffected (T. Kundi\'c 1997,
priv. comm.).  Third, the effects
of ram pressure in groups could be enhanced by the tidal tails produced
during the collision of two galaxies (see Hibbard \etal 1994).  The
intra-group medium will strip the
rarefied gas in the tails more easily than disk gas, perhaps
preventing gas from returning to the merger remnant and thereby halting
any subsequent star formation.
Fourth, we implicitly assume that
that the disk surface densities of fainter
group members are at least as high as for the $L^*$ 
galaxy considered here.  If the
disks of fainter galaxies have substantially lower
restoring force pressure, they may be more affected by ram pressure
stripping.

\subsubsection{The Effects of Viscous Stripping and Thermal Conduction}
Transport processes such as viscous stripping and thermal conduction
are also proposed as gas removal mechanisms in clusters
(Nulsen 1982; Cayatte \etal 1994).  In contrast to ram pressure
stripping, these processes do not depend on the disk's local gravity,
except for the most massive and/or slow moving galaxies.
Therefore, gas is stripped from the galaxy at all radii, not just from the
outermost disk.  In addition, gas transport
is not sensitive to the orientation of a galaxy with respect to
its motion through the intra-group medium, whereas ram pressure
depends on the component of
the galaxy's velocity perpendicular to its disk.  
As in the ram pressure calculation above,
we assume here that there is no replenishment of the gas once it is
swept away and that the galactic gas is uniform.  We also assume
that ram pressure has not first stripped gas from the disk.

The stripping rate due to gas transport is given by
${\dot M_t} \approx \pi {r_d}^2 {\overline {\rho_{IGM}}} \upsilon$,
where $r_d \sim 30$ kpc is the radius of the gas disk
(excluding any ram pressure effects) and 
$\upsilon \approx \sqrt 2 \sigma_r$ 
is the approximate velocity of the galaxy through the intra-group
medium (Nulsen 1982).
For a galaxy
within 300\lith\inv\ kpc of the center of NGC 533,
the gas mass loss rate is then
$\sim 40 h \ \Mdot$ yr$^{-1}$.  Therefore, a typical
galaxy with $\sim 8 \times 10^9 \Mdot$ of HI (NGC 4192 from Cayatte \etal 1994)
could lose almost $100\%$ of its atomic gas in a crossing time
($t_c/t_H = 0.02$ for NGC 533).  However, the HI detected in
many poor group members (Giuricin \etal 1985)
suggests that the effectiveness of this mechanism
is overestimated here.  As pointed out by Kundi\'c \etal (1997), this
approximation neglects the formation of a shock in front
of the galaxy that prevents some fraction of the intra-group medium from
interacting directly with the galactic gas.  In addition, the introduction
of even a small magnetic field term into the transport
equation would limit the effectiveness of the process by
reducing the mean free paths of the plasma particles
(B. Mathews 1997, priv. comm.).  An improved understanding of the
significance of gas transport mechanisms in groups 
awaits detailed simulations.  As in the case of ram pressure, gas
transport may limit the
subsequent star formation of the galaxy, but the requirements
for transforming the galaxy into an earlier morphological type,
{\it e.g.}, the total disruption of the stellar disk or an increase in
the bulge-to-disk ratio, are not part of this picture.

\subsubsection{Implications}
In the preceding discussion, we demonstrated that
the effects of ram pressure stripping are weaker in poor groups
than in rich clusters.  We suggested that,
even if gas disruptive processes 
like ram pressure stripping, viscous stripping, or thermal conduction can
induce star formation as well as truncate it, they are
unlikely to significantly affect the stellar morphology of a galaxy.
Therefore, if one environmental mechanism is responsible
for both the high early type fractions of some 
poor groups and the young stellar
populations of certain early type group galaxies,
galaxy-galaxy encounters provide a possible explanation.
Such interactions can produce bursts of star formation ({\it e.g.},
Londsdale \etal 1984; Kennicutt \etal 1987; Sanders \etal 1988)
in which the gas is consumed or stripped away.

Although our observations of poor groups suggest that the effects of 
cluster environment are not required to produce the
early type fractions and star formation episodes of nearby clusters,
we suspect that the star formation histories
of group and subcluster galaxies will begin to deviate
after the subclump and cluster mix.  Proposed gas
removal processes
including ram pressure stripping, the tidal limitation of galaxy halos,
and galaxy harassment, which are more efficient
in clusters than in groups, may eventually suppress star formation
in cluster galaxies.  Comparative
studies of the HI content of field, group, and cluster galaxies will
help to resolve this issue.

\section{Conclusions}

We use fiber spectroscopy to obtain 963 galaxy redshifts
in the fields of 12 poor groups of galaxies.  When combined
with 39 redshifts from the literature, the survey consists of
1002 galaxy redshifts, 280 of which are group members.  The groups
have mean recessional velocities between 1600 and 7600 \ks.  Nine 
groups, including three Hickson compact groups, are detected
in X-rays by ROSAT (Paper II).
Our conclusions are the following:

1.  Remarkably, there are at least $\sim 20$-50 group
members to absolute magnitudes as faint as
$M_B \sim -14$ to $-16 + 5$log$_{10}$ \lith\ in each of
the X-ray-detected groups, most of
which were previously 
known as groupings of less than five bright ($\simless \Mstar$) 
galaxies ($\S3.1$, Figure 2).  
The large number of group members, the significant early-type populations 
(up to $\sim 55$\% of the membership) and their
concentration in the group centers,
the short crossing times (less than 0.05 of a Hubble time)
through a massive, common halo (see point 2 below),
and the correspondence of the central, giant elliptical with
the optical and X-ray group centroids argue that
the X-ray detected groups are bound systems, not chance superpositions
of unbound galaxies along the line-of-sight, and that the cores of these groups
are close to virialization or virialized.  An exception is
HCG 90, a dynamically evolving 
group whose several core galaxies appear to be
interacting (Longo \etal 1994) 
and whose asymmetric X-ray emission is not coincident with
any of the bright group members.  

Because we increase the membership by a factor of 10 in many of our groups, 
we can, for the first time, determine poor group
velocity dispersions with sufficient precision ($\simless 20\%$) that the
differences among them are statistically meaningful ($\S3.2$).
The velocity dispersions of the X-ray groups range over more
than a factor of two, from
190 \ks\ in HCG 90 and 210 \ks\ in HCG 42 to 430 \ks\ in NGC 741 
and 460 \ks\ in NGC 533.
We are now quantifying the baryonic contribution of
the faint group members which, when 
coupled with our improved group velocity dispersion determinations, 
will allow us to place better constraints on the baryon fraction
of poor groups.

The non-X-ray-detected groups have lower
velocity dispersions ($< 130$ \ks) 
and early-type fractions ($= 0$\%) than the X-ray groups.
We are unable to determine whether any of the three non-X-ray groups 
in our sample are
bound systems with little or no X-ray gas
or whether they are all just superpositions of
unbound galaxies along the line-of-sight.  It is important to note
that the Local Group, a system collapsing for the
first time (cf. Zaritsky 1994), would appear to
have optical properties similar to the non-X-ray groups
if it were moved to their distances.

2.  The group velocity dispersion
does not decrease significantly from the center out to the virial 
radius (typically $\sim 0.5$\lith\inv\ Mpc),
implying that the mass in the group is extended ($\S3.3$, Figure 5).
The mass of the group halo ($\sim 0.5$-$1\times 10^{14}h^{-1} \Mdot$)  
is large compared with that of
the X-ray gas ($\sim 1 \times 10^{12} h^{-5/2} \Mdot$) and of
the galaxies ($\sim 1 \times 10^{13} h^{-1} \ \Mdot$), 
whose individual halos may have been
tidally-limited by the group potential at the time of the
system's formation (Merritt 1984) or subsequently reduced by
collisional stripping (Richstone 1975; Dekel, Lecar, \& Shaham 1980).
The small fraction of the group mass associated with the galaxies
may slow the rate at which they interact
(Governato \etal 1991; Bode 1993; Athanassoula 1997), allowing the group to
virialize before all of its members merge and to exist
as a group for more than a few crossing times.  This picture
is further supported by the evidence, described in point 1 above,
that the cores of some poor groups have survived
long enough to virialize and by the X-ray observations discussed in Paper II.

3.  The giant, brightest elliptical in
each X-ray group (a galaxy that lies within
$\sim 5$-10\lith\inv\ kpc of the X-ray center, Paper II)
occupies a position that
is indistinguishable, in radial velocity and on the sky, from the
center of the group potential ($\S3.4$, Figure 6).
This result suggests that dominant cluster ellipticals,
such as ``cD" galaxies (Matthews, Morgan, \& Schmidt 1965), may form
via the merging of other galaxies
in the centers of poor group-like environments ({\it e.g.,} Hickson 
compact group 90), 
perhaps during the initial
collapse of the group when the conditions for galaxy capture
are most favorable (Merritt 1985).  
The similarity of the
velocity dispersions of the group and the central elliptical
may allow the BGG to subsequently grow through mergers with
bound companions, with slow-moving galaxies on radial orbits,
or with galaxies that fall into the core
via dynamical friction (depending on the group, an $L^*$ galaxy
at a radius of $\sim 250$-400\lith\inv\ kpc would fall to the center in
a Hubble time; see Merritt 1984).
Groups with a central, dominant elliptical may then fall into
richer clusters (Merritt 1985).  This scenario explains
why cD's do not always lie in the spatial and kinematic
center of rich clusters (Zabludoff \etal 1990; 
Dunn 1991; Zabludoff \etal 1993), 
but instead occupy the centers of
subclusters in non-virialized clusters (Geller \& Beers 1983; Bird 1994;
Beers \etal 1995).

4.  The fraction of early-type galaxies in poor groups varies 
significantly ($\S3.5$, Figure 7), 
ranging from that characteristic of the field
($\simless 25$\%) to that more consistent
with rich clusters ($\sim 55$\%).
The high early type fractions of groups like NGC 533, NGC 741,
and HCG 62 are particularly surprising, because
all the 12 sample groups have substantially
lower velocity dispersions (a factor of $\sim 2$-5) and
galaxy number densities (a factor of $\sim 5$-20)
than are typical of 
rich clusters.
What environmental factors might be responsible for the
high early type fractions in these three groups?
First, the effects of disruptive mechanisms
like galaxy harassment (Moore \etal 1996; 
B. Moore 1997, priv. comm.)
on the morphologies of
poor group galaxies are weaker
than in cluster environments.  
Second, although the
tidal limitation of group member
halos,
the expulsion of gas by supernovae-driven
winds, the ram pressure stripping of gas in the outermost disk,
or the removal of gas due to turbulent viscous stripping and thermal conduction
may reduce the galaxies' gas
reservoirs and thus limit their star formation, 
these processes are unlikely to affect galaxy morphology 
({\it e.g.}, it is hard to
transform the bulge-to-disk ratio of a late type spiral into that of an S0
galaxy).  Third, strong gravitational shocks generated
as galaxies pass near the core
on radial orbits might tidally transform
some Sc-Sd galaxies into dwarf ellipticals (B. Moore 1997, priv. comm.),
but only four of the 72 galaxies in our early type sample
are as faint as dE/dSph's 
({\it i.e.,} $M_B \simgreat -16 + 5$log$_{10}$ \lith).
 
In contrast, the kinematics of poor groups make them favorable sites for
galaxy-galaxy encounters (Barnes 1985; Aarseth \& Fall 1980;
Merritt 1985), even though the partial truncation of the galaxy
halos acts to reduce the frequency of such interactions.
Mechanisms such as tidal collisions 
and mergers may influence
the morphologies and star formation histories of some group galaxies.
Another possible explanation of the cluster-like early type fractions
of some poor groups is that
environment is relatively unimportant at late times and that
conditions during the epoch of galaxy formation dominate.

5.  The fraction of early-type group members that have spectral evidence of
on-going or recent star formation in their central regions
(at least $12\%$; $\S3.6$, Figures 8-9) 
is consistent with that seen in rich clusters
with significant substructure ($\sim 15\%$; Caldwell \& Rose 1997).   
If some of the subclusters in these complex
clusters are groups that have
recently fallen into the cluster environment for the first time,
the similarity between the star formation histories of the
early types in the subclusters and of those in our
sample of field groups implies that the cluster environment
and processes such as ram pressure stripping (Gunn \& Gott 1972) are
not required to enhance and/or quench
star formation in these particular
galaxies.  If the recent star formation is tied to the
external environment of the galaxies and not to internal perturbations,
this result is consistent with the picture
in which the star formation histories of some early type 
galaxies are influenced by recent
(within the last 2\lith\inv\ Gyr) galaxy-galaxy encounters.

In future papers, we will discuss the relationships between the X-ray and
optical properties of these groups (Paper II), the
form of the group galaxy luminosity function, and the
baryon fractions of poor groups.

\vskip 0.3in\noindent 
We thank Dennis Zaritsky, Lars Hernquist, and 
Richard Mushotzky for important insights
and detailed comments on the text,
Allan Sandage for help with the art of galaxy classification, 
and Arif Babul, George Blumenthal, Michael Bolte, 
Jack Burns, David Burstein, 
Tomislav Kundi\'c, Bill Mathews, Ben Moore, Julio Navarro,
David Schiminovich, David Spergel, Jacqueline van Gorkom, and 
Steven Zepf for interesting and useful conversations.
AIZ acknowledges support from the Carnegie and Dudley
Observatories, the AAS, NSF grant AST-95-29259, and NASA grant
HF-01087.01-96A.
JM acknowledges support provided by NASA grant NAG 5-2831 and NAG 5-3529.

\vfill\eject
\begin{figure}
\plotone{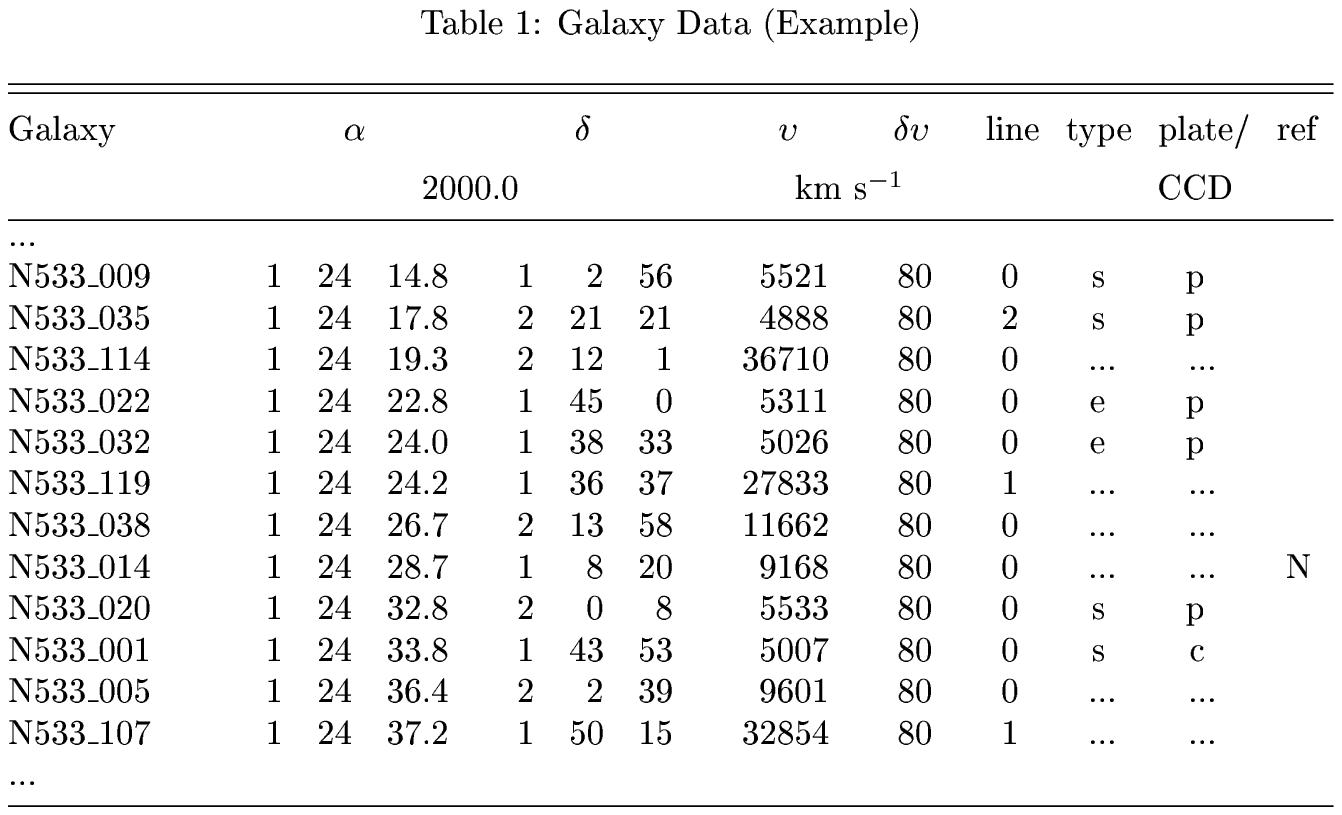}
\end{figure}
\clearpage
\vfill\eject
\begin{figure}
\plotone{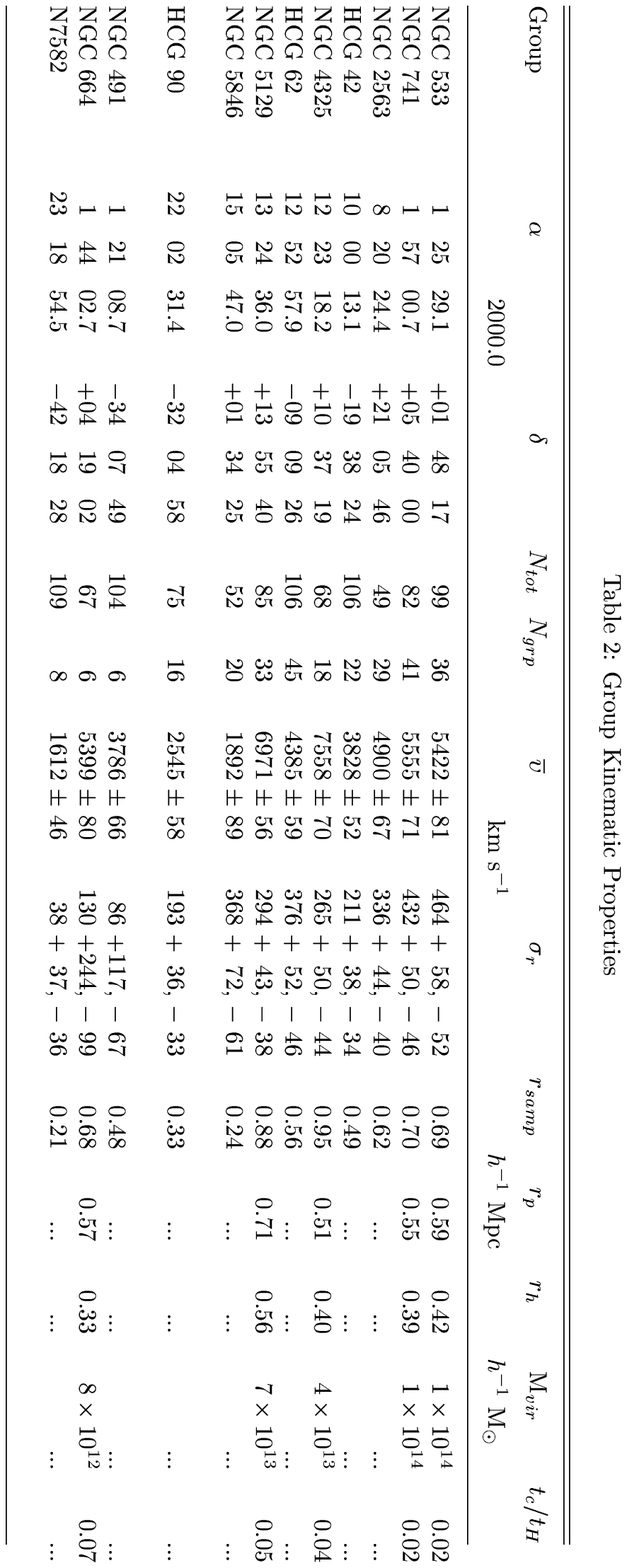}
\end{figure}
\clearpage
\vfill\eject
\centerline{\bf References}
\bigskip
\apj{Aarseth, S.J. \& Fall, S.M. 1980}{236}{43}
\refbook{Athanssoula, E. 1997, preprint.}
\annrev{Bahcall, N.  1977}{15}{505}
\apj{Bahcall, N. 1975}{198}{249}
\nature{Barnes, J. 1989}{338}{123}
\mnras{Barnes, J. 1985}{215}{517}
\aj{Beers, T., Flynn, K., \& Gebhardt, K. 1990}{100}{32}
\aj{Beers, T., Kriessler, J., Bird, C., \& Huchra, J. 1995}{109}{874}
\apj{Beers, T. \& Geller, M. 1983}{274}{491}
\refbook{Binney, J. \& Tremaine, S. 1987, in {\it Galactic Dynamics},
(Princeton: Princeton University Press)}
\aj{Bird, C. 1994}{107}{1637}
\apj{Blumenthal, G., Faber, S., Flores, R., \& Primack, J. 1986}{301}{27}
\apj{Bode, P.W., Berrington, R.C., Cohn, H.N., \& Lugger, P.M. 1994}{433}{479}
\apj{Bode, P.W., Cohn, H.N., \& Lugger, P.M. 1993}{416}{17}
\apj{Bothun, G., Geller, M., Beers, T., \& Huchra, J. 1983}{268}{47}
\apj{Bruzual, A.G., \& Charlot, S. 1993}{405}{538}
\aj{Caldwell, N. \& Rose, J. 1997}{113}{492}
\aj{Caldwell, N. Rose, J.A., Sharples, R.M., Ellis, R.S., \& Bower, 
R.G. 1993}{106}{473}
\aj{Cayatte, V., Kotanyl, C., Balkowski, C., \& van Gorkom, J.H. 1994}{107}{1003}
\aa{Danese, L., de Zotti, G., \& di Tullio, G. 1980}{82}{322}
\apjsup{de Calvalho, R., Ribeiro, A., \& Zepf, S. 1994}{93}{47}
\apj{Dekel, A., Lecar, M., \& Shaham, J. 1980}{241}{946}
\apj{de Oliveira, C.M. \& Hickson, P. 1994}{427}{684}
\aj{Diaferio, A., Geller, M., \& Ramella, M. 1995}{109}{2293} 
\apj{Dressler, A. 1978}{226}{55}
\refbook{Dunn, A. 1991, in {\it Clusters and Superclusters of Galaxies},
Contributed Talks and Poster Papers, NATA Advanced Study Institute, 
Institute of Astronomy, Cambridge, eds. Colless, Babul, Edge, 
Johnstone, \& Raychaudhury, p.13}
\apj{Evrard, A., Metzler, C., \& Navarro, J. 1996}{469}{494}
\apj{Faber, S. \& Jackson, R. 1976}{204}{668}
\mnras{Forbes, D. \& Thomson, R. 1992}{254}{723}
\apj{Giovanelli, R., \& Haynes, M.P. 1985}{292}{404}
\apjlett{Governato, F., Bhatia, R., \& Chincarini, G. 1991}{371}{L15}
\apj{Gunn, J. 1977}{218}{592}
\apj{Gunn, J.E., \& Gott, J.R. 1972}{176}{1}
\refbook{Helou, G., Madore, G., Schmitz, M.,
Bicay, M., Wu, X. \& Bennett, J. 1991, in ``Databases and On-Line Data
in Astronomy," ed. D. Egret \& M. Albrecht (Dordrecht: Kluwer), p. 89.}
\apj{Hernquist, L., Katz, N. \& Weinberg, D. 1995}{442}{57}
\aj{Hibbard, J.E., Guhathakurta, P., van Gorkom, J.H., \& Schweizer, F. 1994}{107}{67}
\apj{Hickson, P., de Oliveira, M., Huchra, J., \& Palumbo, G. 1992}{399}{353}
\apj{Hickson, P., Huchra, J., \& Kindl, E. 1988}{331}{64}
\apj{Hickson, P. 1982}{255}{382}
\apjsup{Huchra, J.P., Geller, M.J., \& Corwin, H. 1995}{99}{391}
\apj{Hughes, J. 1989}{337}{21}
\apj{Hunsberger, S., Charlton, J., \& Zaritsky, D. 1996}{462}{50}
\aj{Jarvis, J.F. \& Tyson, J.A. 1981}{86}{476}
\apj{Kennicutt, R.C., Roettiger, K.A., Keel, W.C., Van der Hulst, J.M.
\& Hummel, E. 1987}{93}{1011}
\apjsup{Kennicutt, R.C. 1992a}{79}{255}
\apj{Kennicutt, R.C. 1992b}{388}{310}
\apjlett{King, I. 1972}{174}{L123}
\aj{King, I. 1966}{71}{64}
\aj{Kirshner, R., Oemler, A., Schechter, P., \& Shectman, S. 1983}{88}{1285}
\apj{Kormendy, J. \& Bender, R. 1996}{464}{119}
\refbook{Kundi\'c, T., Hernquist, L., \& Spergel, D. 1997, preprint}
\pasp{Larson, R.B. \& Dinerstein, H.L. 1975}{87}{911}
\aj{Leonardi, A. \& Rose, J. 1996}{111}{182}
\refbook{Lin, H. 1995, Ph.D. Thesis, Harvard University}
\apj{Lonsdale, C.J., Persson, S.E., \& Matthews, K. 1984}{287}{95}
\aa{Longo, G., Busarello, G., Lorrenz, H., Richter, G., \& Zaggia, S. 1994}
{282}{418}
\mnras{Maddox, J., Efstathiou, G.; Sutherland, W. J.; Loveday, J. 1990}{243}{692}
\refbook{Mamon, G.A. 1992, in {\it Physics of Nearby Galaxies:  Nature or
Nurture?}, ed. T.X. Thuan, C. Balkowski, \& J. Tran Thanh Van, 12th Moriond
Astrophysics Meeting (Gif-sur-Yvette:  Editions Fronti\`eres), p.367}
\refbook{Matthews, T.A., Morgan, W.W., \& Schmidt, M. 1965, 
in {\it Quasi-Stellar Sources and Gravitational Collapse}, eds. Robinson, 
Schild, and Schucking (Chicago: University of Chicago Press), p.105}
\apjsup{McElroy, D.B. 1995}{100}{105}
\aj{Merritt, D. \& Tremblay, B. 1996}{111}{2462}
\apj{Merritt, D. 1985}{289}{18}
\apj{Merritt, D. 1984}{276}{26}
\refbook{Mink, D. J. \& Wyatt, W. F. 1995, Astronomical Data Analysis 
Software and Systems IV, ASP Conference Series, Vol. 77,
1995, R.A. Shaw, H.E. Payne, and J.J.E. Hayes, eds., p. 496.}
\nature{Moore, B., Katz, N., Lake, G., Dressler, A., \& Oemler, A. 1996}{379}{613}
\refbook{Mulchaey, J. S. \& Zabludoff, A.I. 1997, ApJ, submitted. (Paper II)}
\apjlett{Mulchaey, J. S., Davis, D. S., Mushotzky, R. F. \& Burstein, D. 1993}
{404}{L9}
\apjlett{Mulchaey, J. S., Mushotzky, R. F., Burstein, D. \& Davis, D. S. 1996a}
{456}{L5}
\apj{Mulchaey, J., Davis, D., Mushotzky, R., \& Burstein, D. 1996b}{456}{80}
\apj{Navarro, J.F., Frenk, C.S., \& White, S.D.M. 1996}{462}{563}
\refbook{Oemler, A. 1992, in {\it Clusters and Superclusters of Galaxies},
ed. A.C. Fabian (Dordrecht: Kluwer), p.29}
\mnras{Nulsen, P. 1982}{198}{1007}
\apj{Ostriker, J., Lubin, L., \& Hernquist, L. 1995}{444}{61}
\aj{Peebles, P.J.E. 1970}{75}{13}
\apj{Pildis, R., Bregman, J., \& Evrard, A. 1995}{443}{514}
\mnras{Ponman, T., Bourner, P., Ebeling, H., \& Bohringer, H. 1996}{283}{690}
\nature{Ponman, T. \& Bertram, D. 1993}{363}{L51}
\apj{Ramella, M., Geller, M.J., \& Huchra, J.P. 1989}{344}{57}
\apj{Richstone, D. \& Malumuth, E. 1983}{268}{30}
\apj{Richstone, D. 1975}{200}{535}
\apj{Rose, J.A. 1977}{211}{311}
\apjsup{Rubin , V., Hunter, D., \& Ford, W. 1991}{76}{153}
\apj{Sanders, D.B., Soifer, B.T., Elias, J.H., Matthews, K., 
\& Madore B.F. 1988}{325}{74}
\apj{Schechter, P.L. 1976}{203}{297}
\apj{Schombert, J. 1988}{328}{475}
\aj{Schweizer, F. \& Seitzer, P. 1992}{104}{1039}
\science{Schweizer, F. 1986}{231}{227}
\refbook{Shectman, S.A., Schechter, P.L., Oemler, A.A., Tucker, D.,
Kirshner, R.P., \& Lin, H. 1992, in Clusters and Superclusters
of Galaxies (ed. Fabian, A.C.) 351-363 (Kluwer, Dordrecht)}
\apj{Shectman, S.A., Landy, S.D., Oemler, A., Tucker, D.L., Lin, H.,
Kirshner, R.P.; Schechter, P.L. 1996}{470}{172}
\refbook{Trager, S.C. 1997, Ph.D. Thesis, 
University of California, Santa Cruz.}
\refbook{Tremaine, S. 1990, in {\it Dynamics and Interactions of
Galaxies}, ed. Wielen (Berlin: Springer-Verlag), p.394}
\apj{Tonry, J. \& Davis, M. 1981}{246}{680}
\aa{van den Bergh, S. 1992}{264}{75}
\aa{van der Kruit, P. \& Searle, L. 1981}{95}{105}
\apj{Weil, M. \& Hernquist, L. 1996}{460}{101}
\mnras{White, S., Briel, U., \& Henry, P. 1993}{261}{L8}
\apj{Whitmore, B., Gilmore, D., \& Jones, C. 1993}{407}{489}
\apj{Yahil, A. \& Vidal, N. 1977}{214}{347}
\apj{Yun, M.S., Verdes-Montenegro, L., del Olmo, A., Perea, J. 1997}{475}{21}
\apjlett{Zabludoff, A.I. \& Zaritsky, D. 1995}{447}{L21}
\aj{Zabludoff, A.I., Huchra, J.P., Geller, M.J., \& Vogeley, M.S. 1993}{106}{1273}
\apjsup{Zabludoff, A.I., Huchra, J.P., \& Geller, M.J. 1990}{74}{1}
\apj{Zaritsky, D., Smith, R., Frenk, C., \& White, S. 1997}{478}{39}
\aj{Zaritsky, D., Zabludoff, A.I., \& Willick, J. 1995}{110}{1602}
\apj{Zaritsky, D. \& White, S. 1994}{435}{599}
\refbook{Zaritsky, D. 1994, in {\it The Local Group:  Comparative and Global
Properties}, eds. A. Layden, R. C. Smith, and J. Storm, ESO Conference and
Workshop Proceedings No. 51, p.187}
\apj{Zepf, S. \& Whitmore, B. 1993}{418}{72}

\vfill\eject\noindent

\centerline{\bf Figure Captions}

\bigskip
\noindent
{\bf Figure 1:} The residuals of the heliocentric velocities from this
paper and HI velocities from the literature
(NASA Extragalactic Database (NED), Helou \etal 1991)  
vs. our internal velocity
error estimates for 39 galaxies.  The mean residual of 13 \ks\ (solid line) 
is small
compared with the {\it rms} deviation of the residuals (83 \ks) and is
consistent with
the mean velocity of 94 stars ($-17$ \ks) 
that were serendipitously observed with the
same instrument (dashed line).  Therefore, we do not apply a zero-point
correction to the galaxy velocities.  We adopt 80 \ks\ as
an estimate of the true velocity error in cases where the internal error
is smaller than 80 \ks.  This error estimate is consistent with
the average external error estimate of $\sim$70 \ks\ for 
the Las Campanas Redshift Survey (Shectman \etal 1996), which employed the same
fiber spectrograph setup.

\noindent
{\bf Figure 2:} Galaxy velocity distributions from
0 to 30000 \ks\ for the 12 poor groups
in our sample.  The first nine groups are detected by the ROSAT
PSPC, the last three are not (Paper II).  Note that
other groups or richer clusters are detected in many of the group fields. 
The shaded histograms indicate
the N$_{grp}$ group members identified with the $3\sigma$-clipping algorithm
described in $\S3.1$.  Because each field is sampled
to a different absolute magnitude limit and
out to a different physical radius, the histograms do not precisely
reflect the actual differences in group galaxy number densities (see
$\S3.1$ and Figure 4).  The histograms do reveal large
populations of group members down to absolute magnitudes 
of $M_B \sim -14$ to $-16 +5$log$_{10}$ \lith.

\noindent
{\bf Figure 3:} The projected spatial distributions of the members
of the 12 sample groups.  The total angular size of each plot is
$1.62 \times 1.62$ degrees, slightly larger than
the fiber spectrograph field of
$1.5\times 1.5$ degrees.  Digitized scans of Palomar Sky Survey
or UK Schmidt plates (APM Survey (Maddox \etal 1990);
STScI Digitized Sky Survey) are not complete for every group;
hence, we observe no galaxies in the westernmost fifth of
the NGC 741 field, in the northernmost fifth of the NGC 5129 field,
in the northernmost fifth of the HCG 90 field, and the northernmost tenth
of the NGC 7582 field (as indicated by the dashed
lines).  The galaxies are typed to an apparent
magnitude of $m_B \sim 17$.
Early type galaxies (E, S0) are denoted by ``0", late
type galaxies (non-E, non-S0) by ``S".  The filled circles mark the
untyped group members.  The scale bar below each group name
is 0.3\lith\inv\ Mpc.  The early types tend to be more
concentrated in the group centers than the late types.

\noindent
{\bf Figure 4:} The number of group galaxies within
a projected radius of 0.3\lith\inv\ Mpc and with
absolute magnitudes of $M_B \simless -17 + 5$log$_{10}$ \lith.
The observed number counts of members within these limits are shaded.
To roughly compensate for incomplete sampling down to the magnitude
limit (completeness is indicated
by the fraction above each histogram bar), we assume that
the fraction of all unobserved galaxies that are group members 
is the same as the fraction of all observed galaxies that are members.
The white histogram shows these ``corrected" group galaxy counts.
Magnitude errors introduce more uncertainty into
the ``corrected" galaxy counts (see $\S3.1$), so a
small difference between the counts of two groups is not significant.
Nevertheless, this plot shows that the trend is for
non-X-ray-detected groups
to have lower galaxy densities than are typical of X-ray groups.
For the same radial and magnitude cuts, the Coma cluster (NED)
has 83 members (a lower limit because we make no correction
for incompleteness), implying that its
galaxy density is $\sim 5$-$20\times$ that of
the poor groups.

\noindent
{\bf Figure 5:}  The velocity offset vs. the projected radial offset
of 204 X-ray group members from the central, giant elliptical (BGG)
in each group.  The velocity offset is normalized with the internal
velocity dispersion of the BGG ($\sigma_{BGG}$).  
We plot only the seven X-ray groups for which $\sigma_{BGG}$ 
is known (HCG 42, HCG 90, NGC 2563, NGC 5129, NGC 5846, NGC 533, NGC 741).
We calculate the normalized velocity dispersion ($\delta$)
(i.e., the {\it rms} deviation in the normalized velocity offset) 
within each radial bin (dashed lines).  
We show the magnitude of $\delta$ as a scale
bar at the bottom of each bin.  The first bin ($0 < r \leq 100$\lith\inv\ kpc; 
47 galaxies) has $\delta = 1.3$, the 
second ($100 < r \leq 200$\lith\inv\ kpc; 40
galaxies) has $\delta = 1.4$,
the third ($200 < r \leq 300$\lith\inv\ kpc; 34 galaxies) has 
$\delta = 1.4$, the fourth ($300 < r \leq 500$\lith\inv\ kpc; 43 galaxies) has 
$\delta = 1.3$, and the fifth 
($500 < r \leq 1000$\lith\inv\ kpc; 40 galaxies) has $\delta = 1.3$.
The differences in $\delta$ with radius are not statistically significant.
The flatness of the $\delta$ profile
to at least $\sim 0.5$\lith\inv\ Mpc, in contrast to the
factor of two decrease expected for a Keplerian system,
suggests that the group galaxies move through a common dark halo.
In all bins, $\delta > 1$, which shows that matter is 
dynamically cooler in the BGG than in the surrounding group.

\noindent
{\bf Figure 6:}  
(a) The kinematic ($y$) and projected spatial ($x$) offsets of the
central, giant elliptical (filled squares) and other group
members (filled circles)
from the group centroid for the nine X-ray groups.  
The velocity offset
of each galaxy from the mean velocity of the group
is divided by the group velocity dispersion \sigmar\ to compensate for
differences in the size of the group global potentials.
The $y$ errorbars
represent the $68\%$ confidence limits on $y$ obtained
from adding the errors in the group mean velocity and the BGG velocity
in quadrature.  To estimate the $68\%$ error in $x$ 
for a BGG in a group with $N_{grp}$ members, we
use a statistical jackknife test in which samples of $N_{grp}$ galaxies
are drawn from HCG 62, the group with the most members.
For each BGG,
we adopt the $rms$ deviation in the distribution of $x$ ($\epsilon_x$)
for these samples as the $x$ error.
(For the BGG in HCG 62, we use the smallest $x$ error calculated
among the other groups).
Note that the incomplete, asymmetric 
spatial sampling of the fiber field in NGC 741 and
NGC 5129 (see Figure 3)
contribute to their high $x$ values and that their $x$ errors
are consequently underestimated here.
(b) The distributions of 
the statistic $R$ (see text) for all of the group members (solid) 
and for the subset of BGGs (shaded).
A galaxy that has a large peculiar motion
and/or that lies outside the projected group core has a larger $R$ value
than a galaxy at rest in the center of the group potential.
A Student's t-test gives $< 3 \times 10^{-5}$ as the probability that
the means of the non-BGG and BGG distributions are consistent.
The heavy line shows the distribution of
$R$ obtained by assuming that all of the BGG's lie in the centers of 
their groups.  A t-test fails to differentiate
between the means of the model $R$ distribution and that of the BGGs (shaded)
at better than the $95\%$ confidence level.
We conclude that (1) the BGG's are significantly 
more concentrated than the other group members and (2)
to within the errors, the BGG's occupy the center of the group potential.

\noindent
{\bf Figure 7:}  Histogram of the early type (E, S0) fractions of 
the 12 sample groups.  Note that the $\sim$55$\%$
fractions for three X-ray groups,
HCG 62, NGC 741, and NGC 533, are consistent with the early type fractions
typical for rich clusters over similar
radii (0.5-1.0\lith\inv\ Mpc; Whitmore, Gilmore, \& Jones 1993).  The
early type fractions in these three poor
groups are inconsistent with those of groups like NGC 2563, which have 
morphological populations characteristic of the field
($\sim$30$\%$; Oemler 1992).  Of the 6-8 group members found in 
the three non-X-ray groups (shaded), we find no early types.
The inset shows the correlation between early type fraction $f$ and
group velocity dispersion \sigmar\ for the 12 sample groups.  The solid
line is an unweighted fit to the data, the dashed line
is a fit weighted by the velocity dispersion 
errors.  In both
cases, the correlation is significant at the $>0.999$ level, implying
either that galaxy morphology is set by the local potential size at the
time of galaxy formation (Hickson, Huchra, \& Kindl 1988) and/or that 
\sigmar\ and $f$ increase as a group evolves (Diaferio \etal 1995).

\noindent
{\bf Figure 8:} (a)  Observed, unfluxed spectra of
the central regions of four early type group
members with significant [OII]
emission ($> 5$\AA).  The spectra are smoothed to $1.5\times$ the instrument
resolution of 5-6\AA.  The scale bar to the left of each
spectrum is 50 counts.  (b)  Spectra of the central regions
of seven early type group members for which the
Ca II H+$H\epsilon$ line is stronger than 
the Ca II K line.  The spectrum of N$741\_{020}$ in (a)
and the post-star formation spectra in
(b) comprise $12\%$ of the sample of 64 early type group members
for which we have spectra.  This
fraction of star forming and post-star forming early type
galaxies is consistent
with that typical of rich clusters with significant substructure
($\sim 15\%$; Caldwell \& Rose 1997).  

\noindent
{\bf Figure 9:} Distributions of the 4000\AA\ break $D_{4000}$ for 
the early type group members (white) and for 
the subset of eight star forming and post-star forming galaxies
in Figure 8 (shaded).
The subsample clearly consists of bluer galaxies
(the blue galaxy not included in the subsample is an AGN, H$90\_017$). 
The combination of small $D_{4000}$
and large Ca II H+$H\epsilon$ to Ca II K ratio
indicates that the
subsample galaxies have younger stellar populations than are typical
for early type galaxies in poor groups.
(Note that this distribution of 
$D_{4000}$ is not directly comparable to that of the Las Campanas
Redshift Survey (LCRS; Zabludoff \etal 1996), because
the falling detector response in the blue for our lower redshift group sample 
causes a steepening (reddening) of the 4000\AA\ break
relative to the more distant field sample drawn from the LCRS.)

\eject
\begin{figure}
\plotone{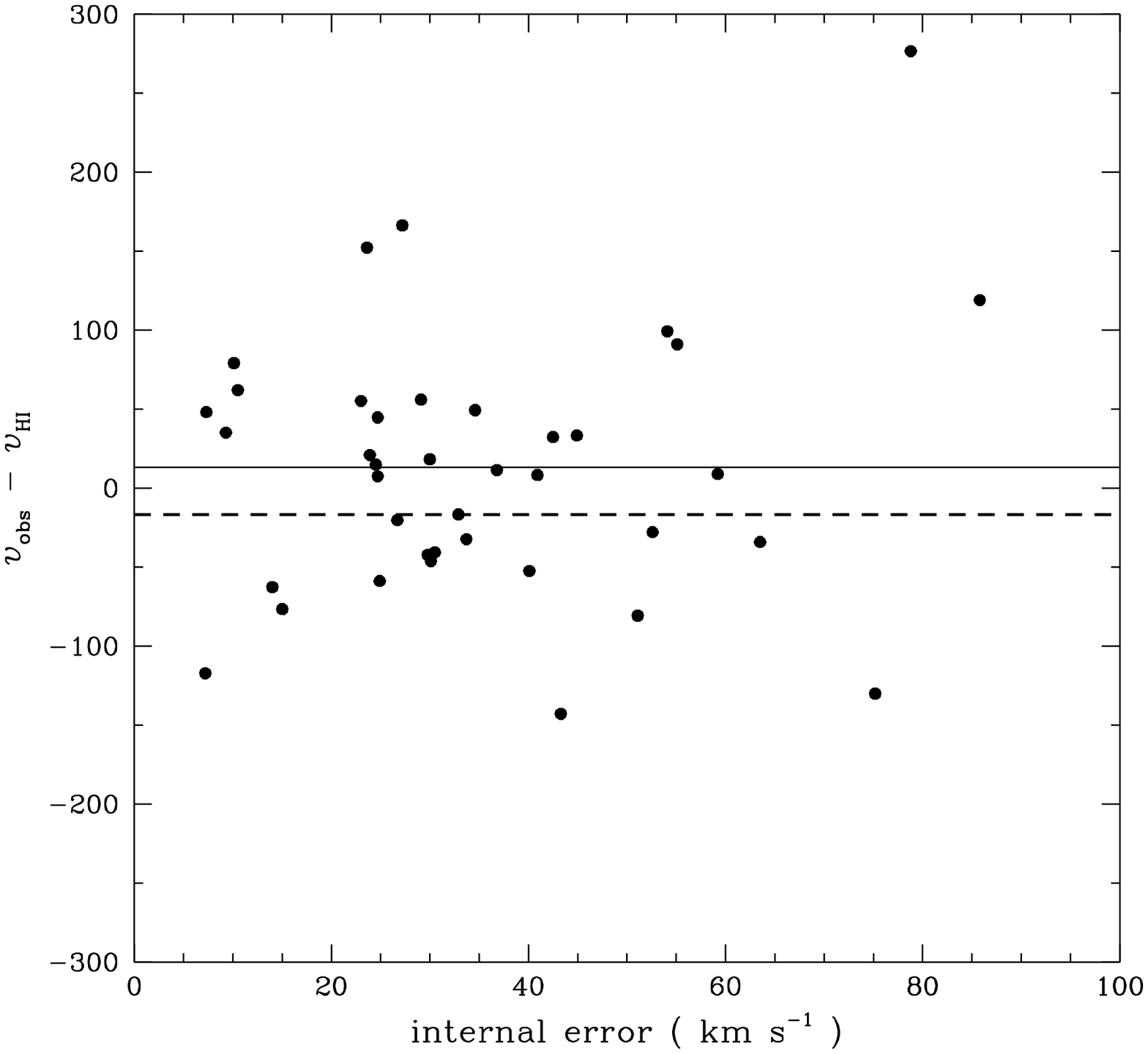}
\caption{}
\end{figure}
\clearpage
\vfill\eject
\clearpage
\begin{figure}
\plotone{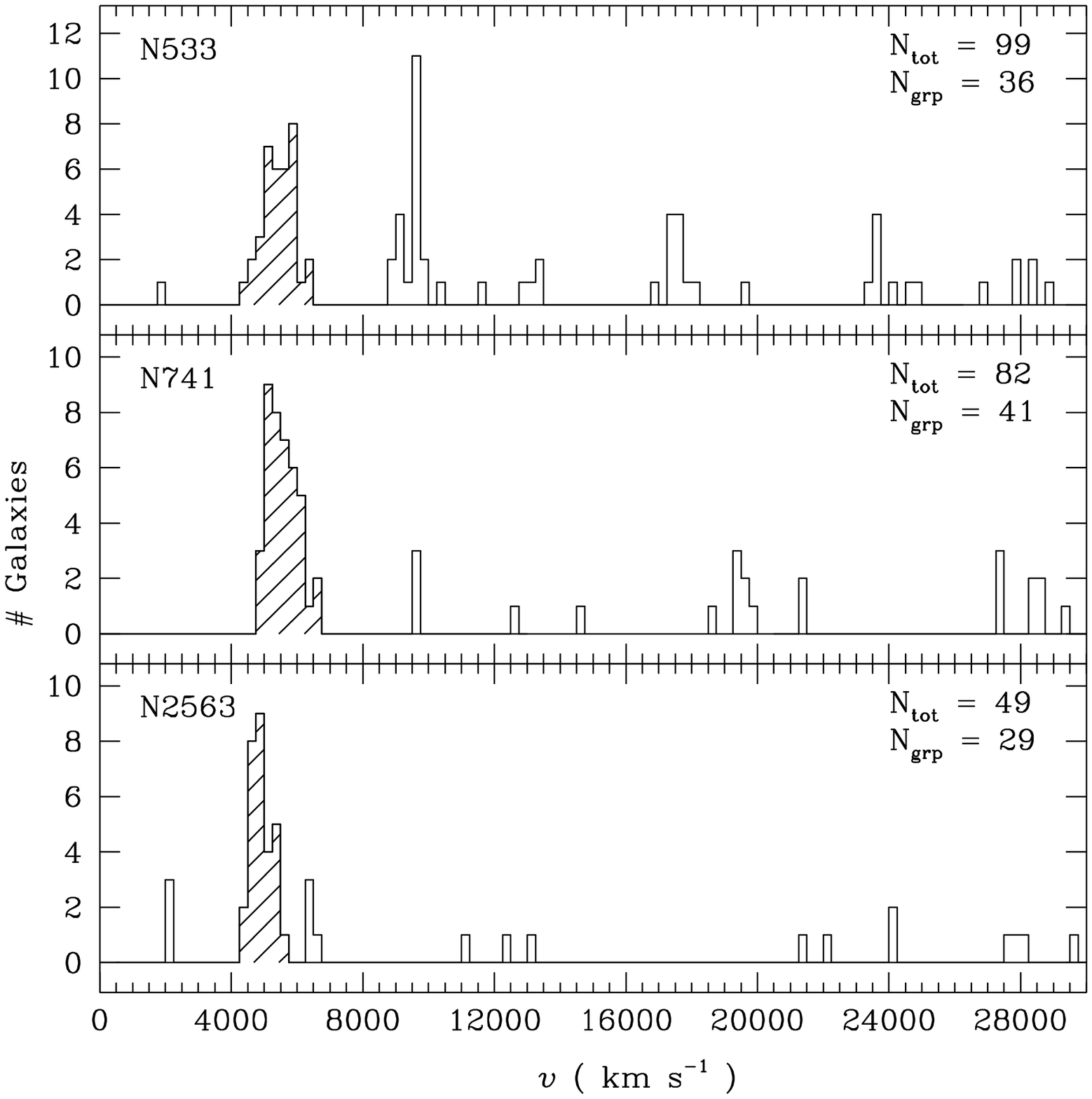}
\caption{}
\end{figure}
\clearpage
\vfill\eject
\clearpage
\setcounter{figure}{1}
\begin{figure}
\plotone{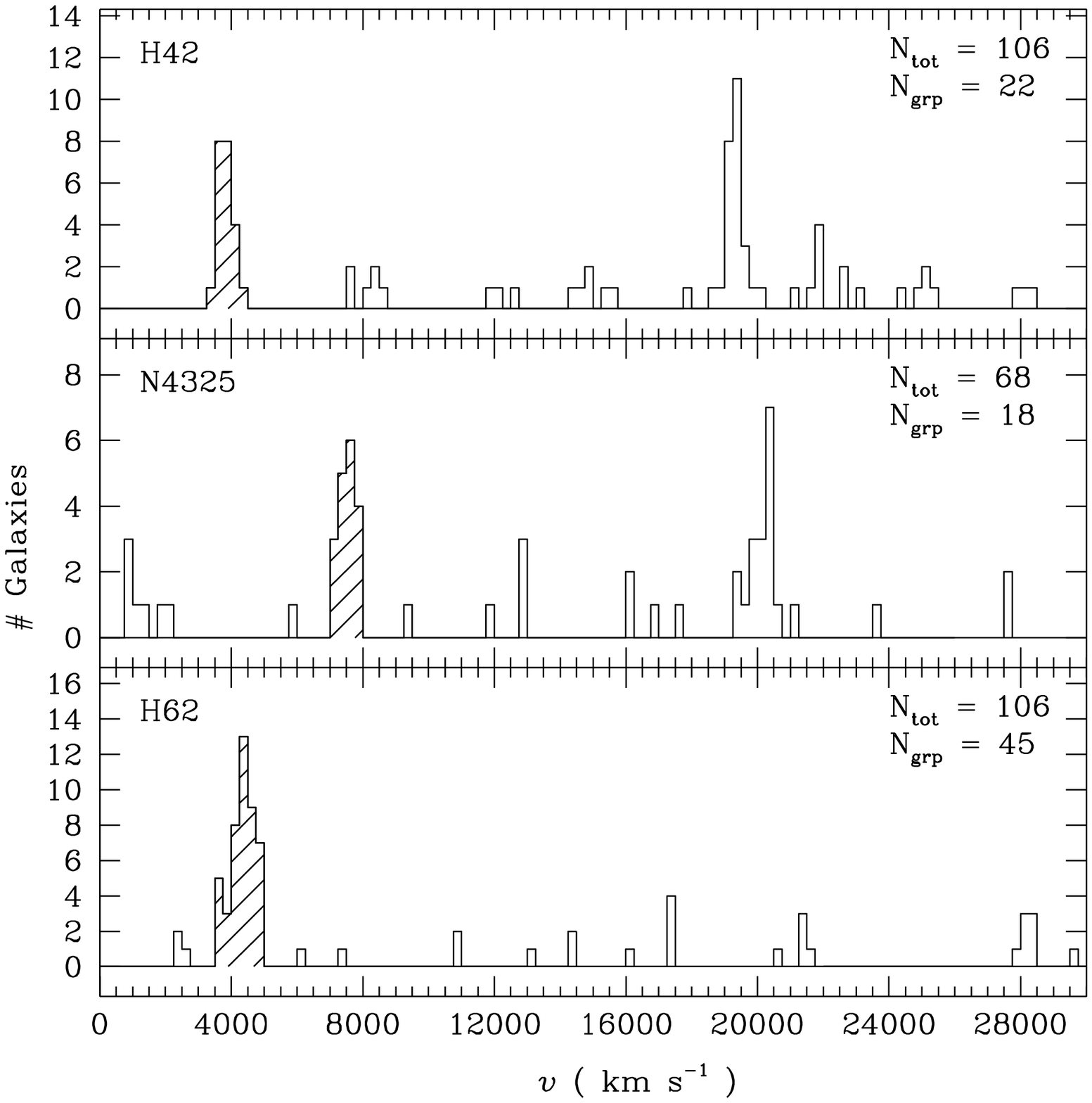}
\caption{{\it cont. }}
\end{figure}
\clearpage
\vfill\eject
\clearpage
\setcounter{figure}{1}
\begin{figure}
\plotone{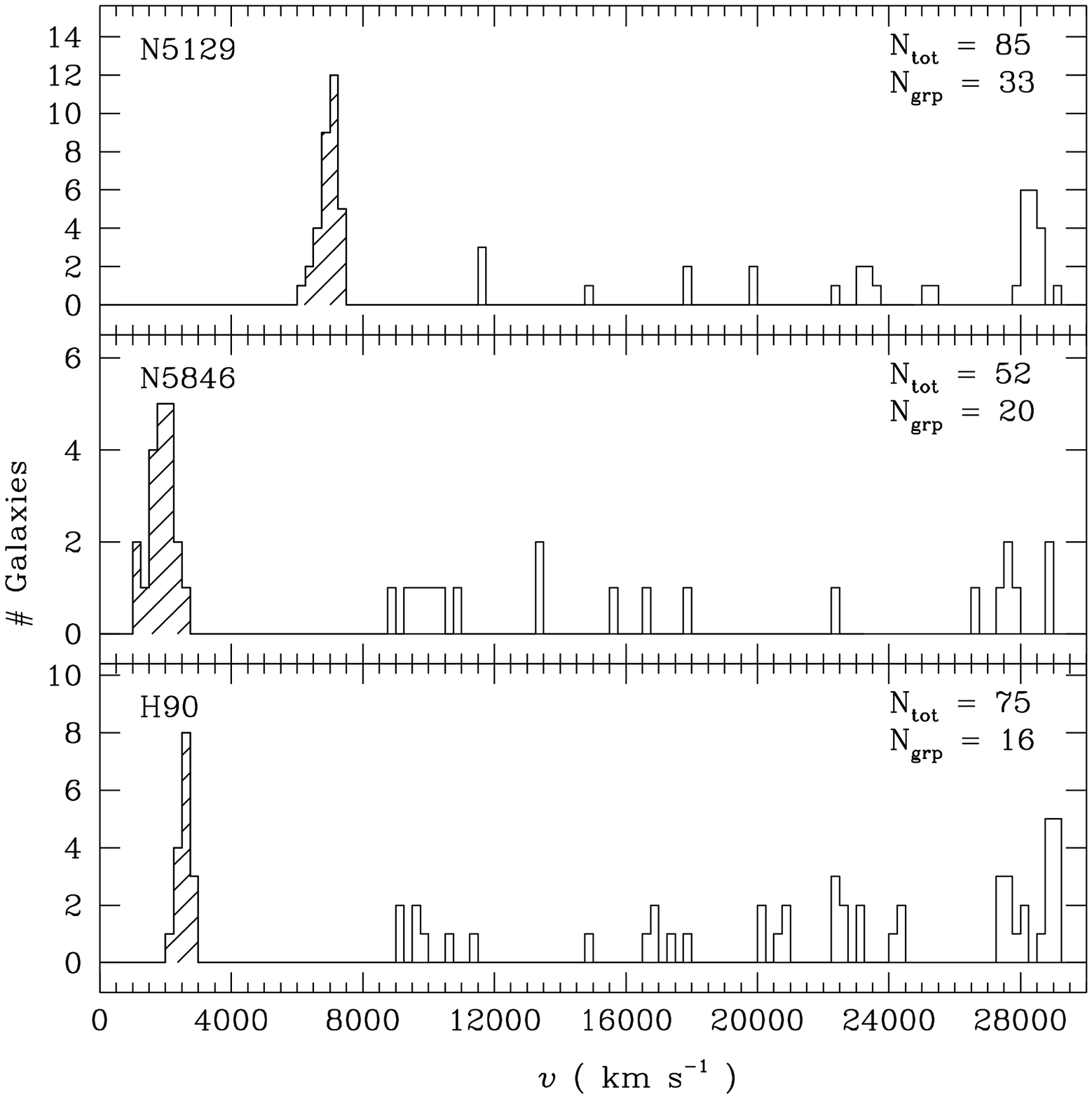}
\caption{{\it cont. }}
\end{figure}
\clearpage
\vfill\eject
\clearpage
\setcounter{figure}{1}
\begin{figure}
\plotone{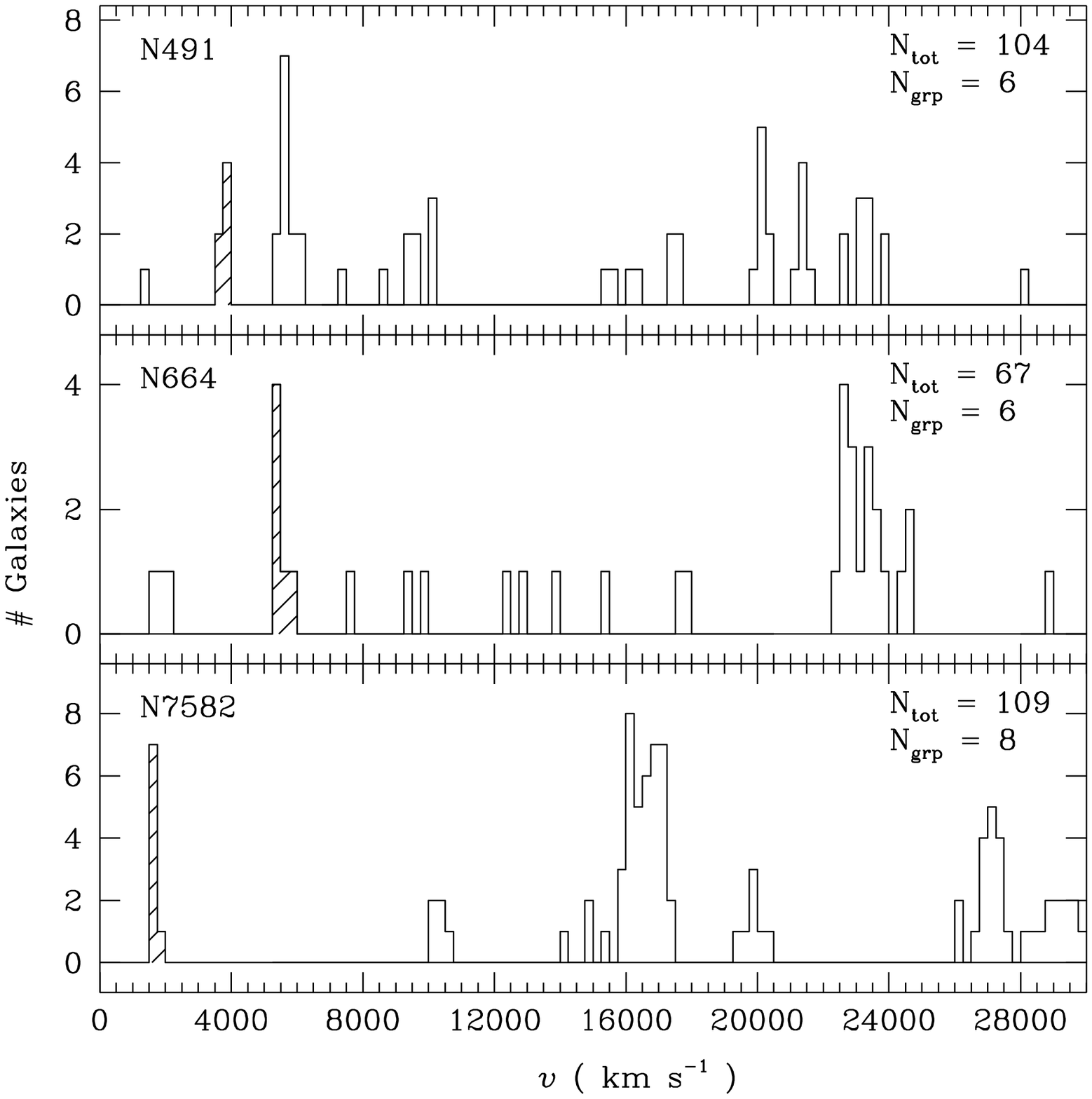}
\caption{{\it cont. }}
\end{figure}
\clearpage
\vfill\eject
\clearpage
\setcounter{figure}{2}
\begin{figure}
\plotone{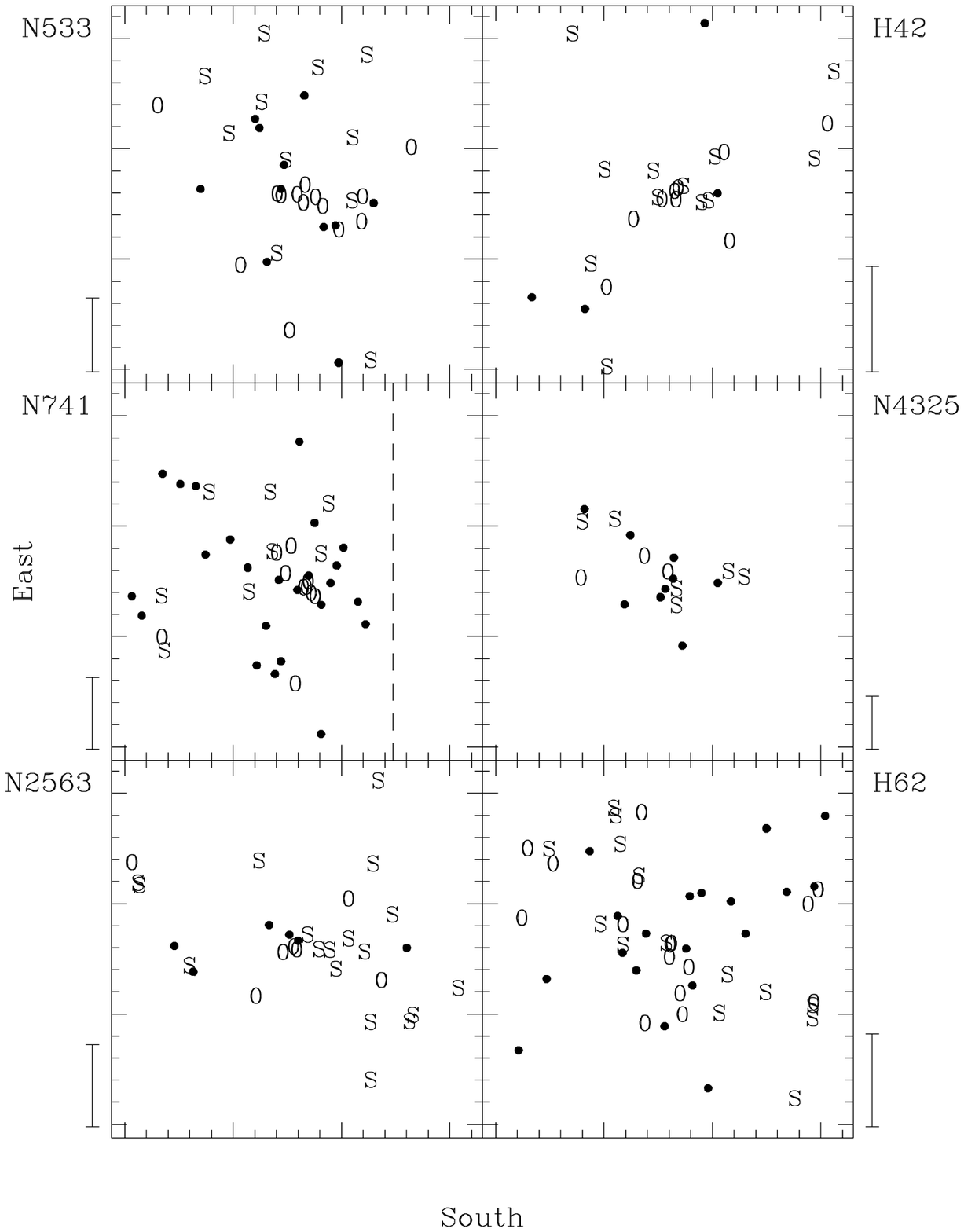}
\caption{}
\end{figure}
\clearpage
\vfill\eject
\clearpage
\setcounter{figure}{2}
\begin{figure}
\plotone{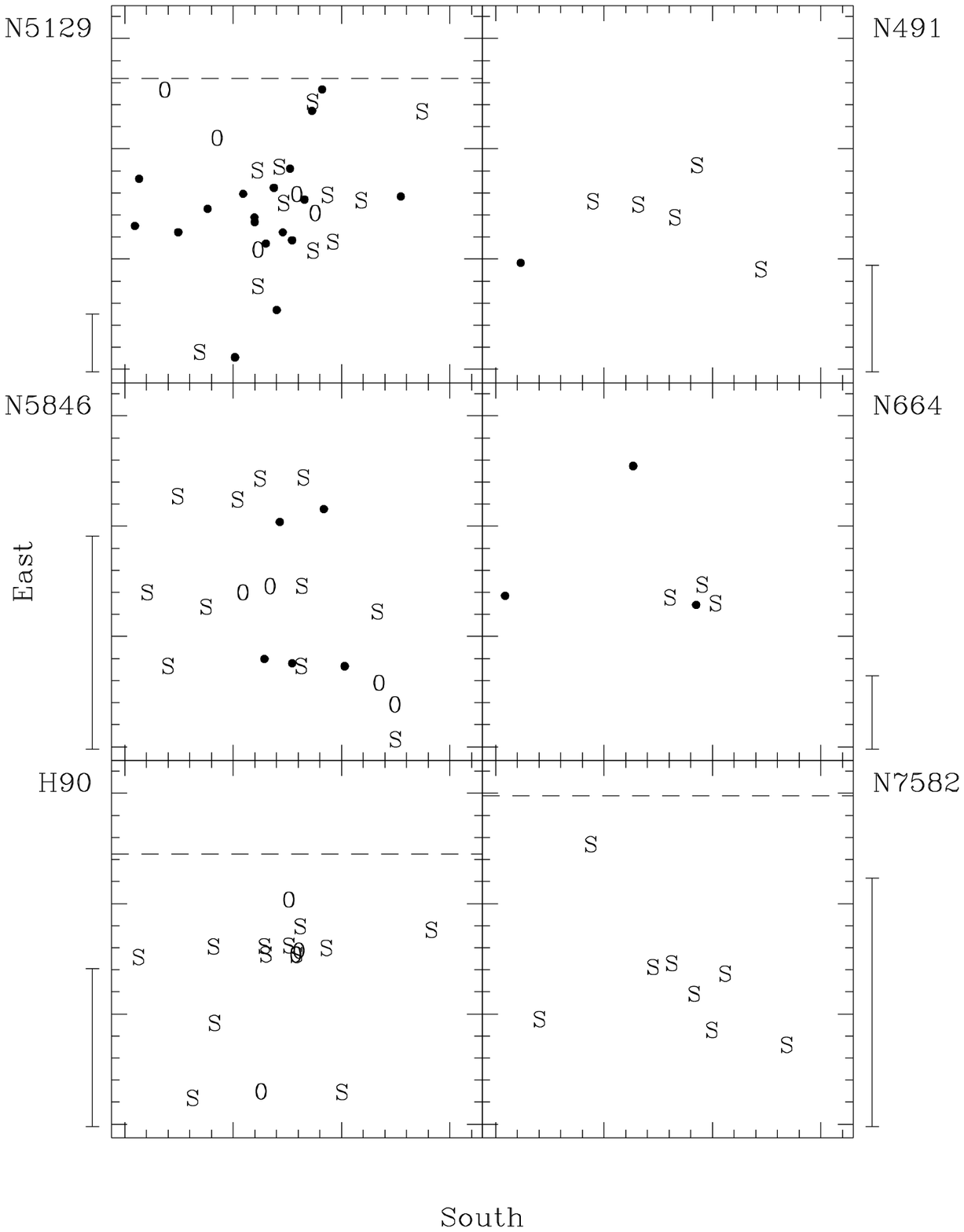}
\caption{{\it cont. }}
\end{figure}
\clearpage
\vfill\eject
\clearpage
\setcounter{figure}{3}
\begin{figure}
\plotone{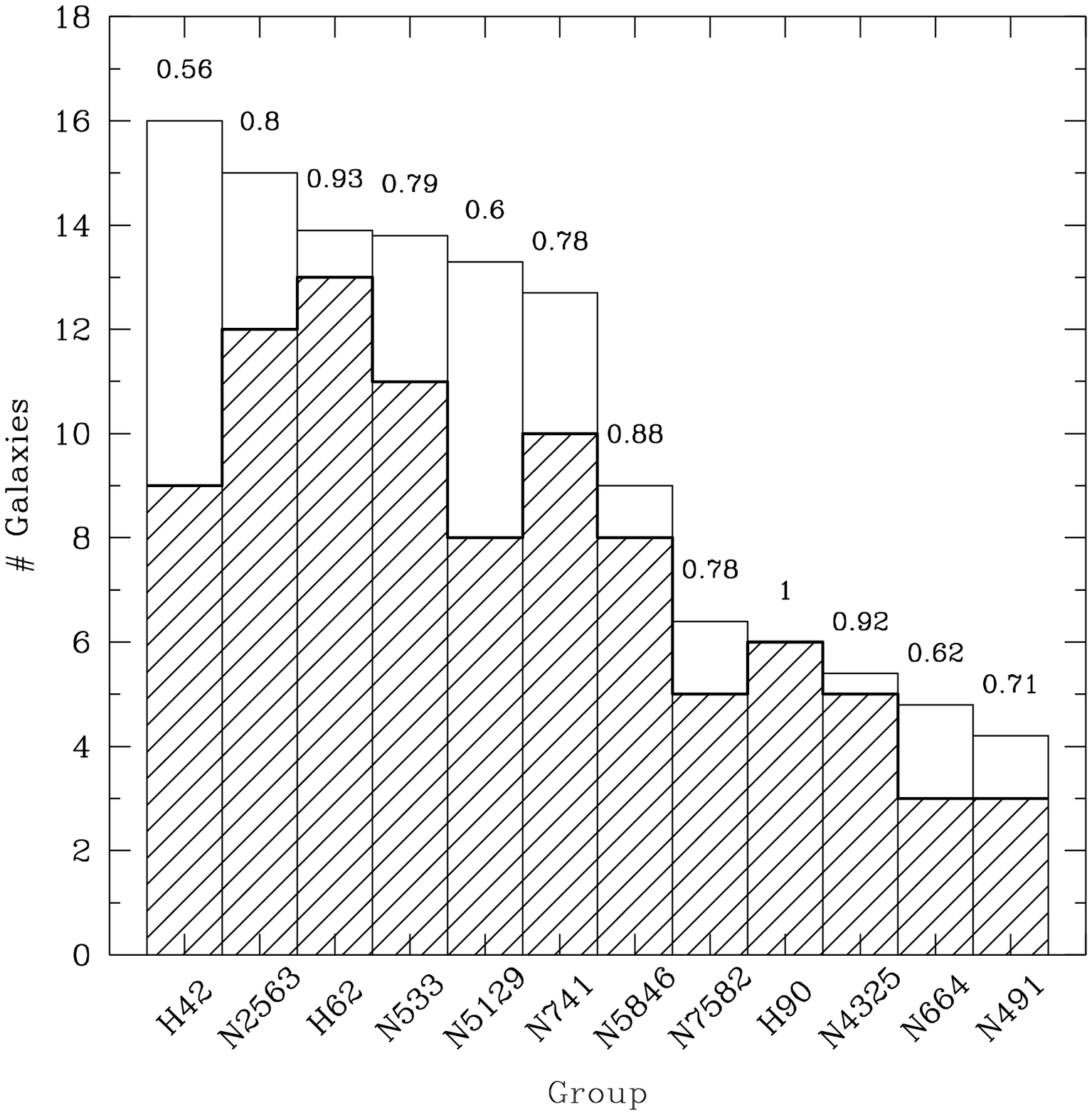}
\caption{}
\end{figure}
\clearpage
\vfill\eject
\clearpage
\begin{figure}
\plotone{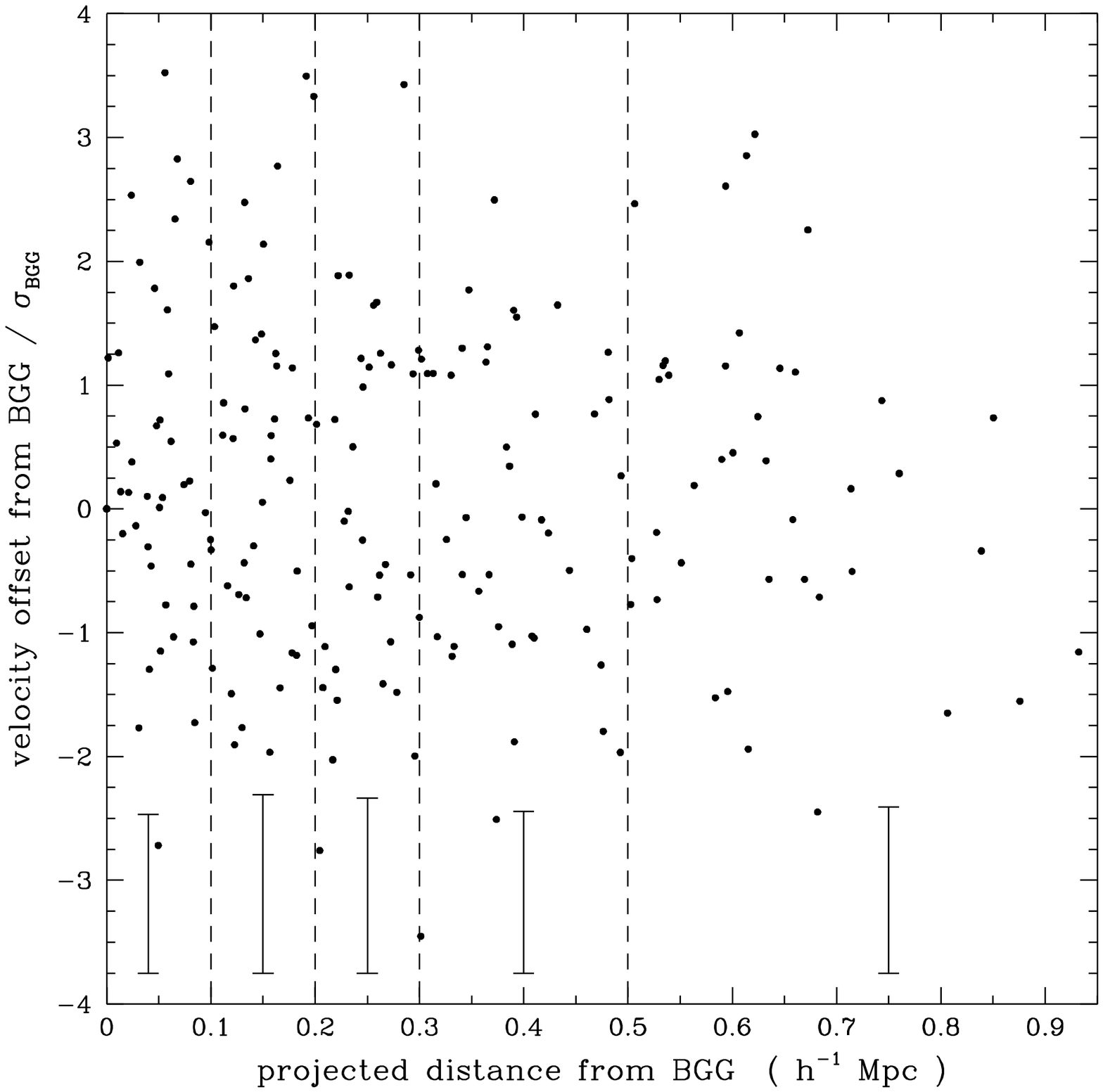}
\caption{}
\end{figure}
\clearpage
\vfill\eject
\clearpage
\begin{figure}
\plotone{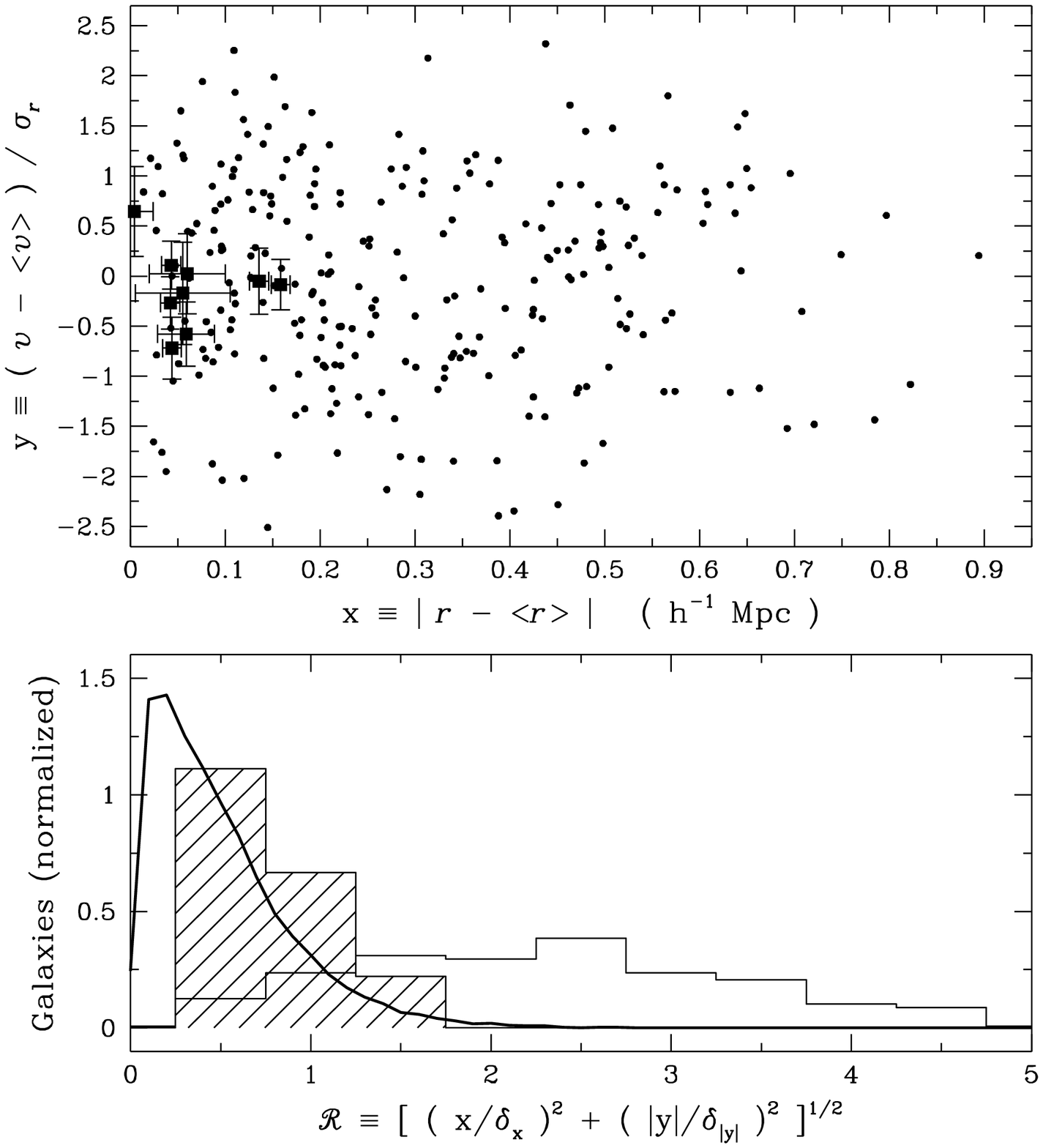}
\caption{}
\end{figure}
\clearpage
\vfill\eject
\clearpage
\begin{figure}
\plotone{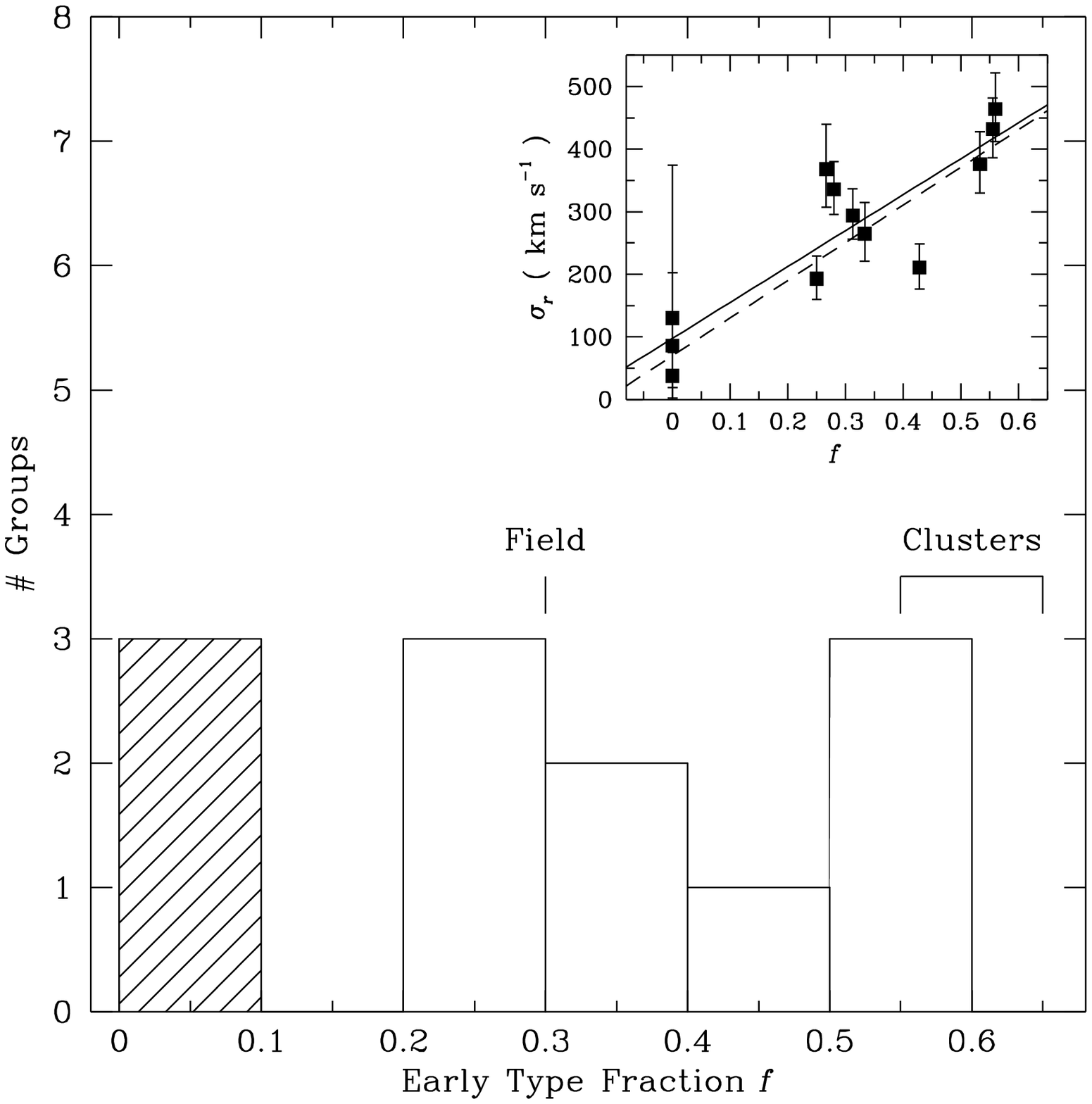}
\caption{}
\end{figure}
\clearpage
\vfill\eject
\clearpage
\begin{figure}
\plotone{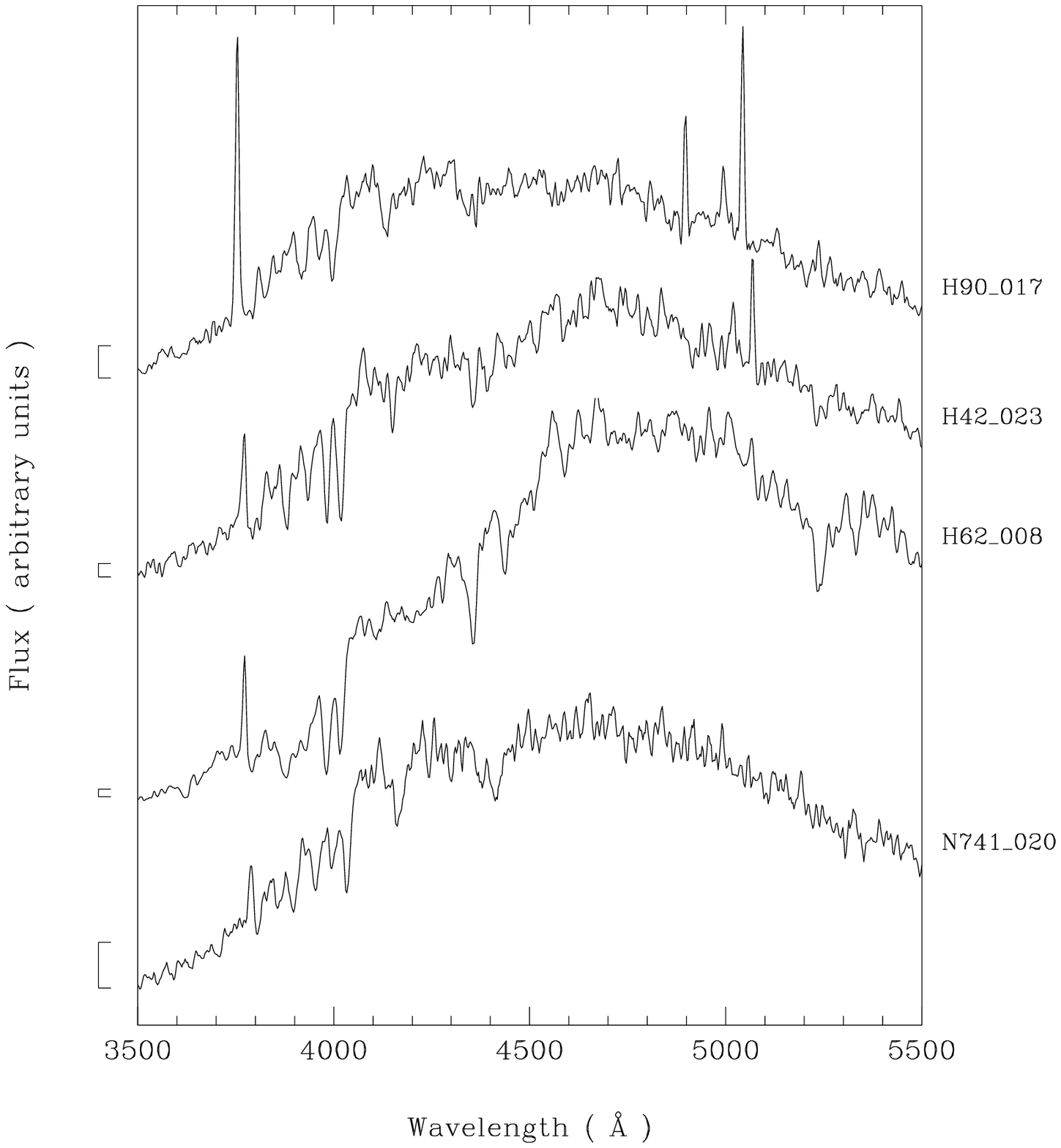}
\caption{}
\end{figure}
\clearpage
\vfill\eject
\clearpage
\setcounter{figure}{7}
\begin{figure}
\plotone{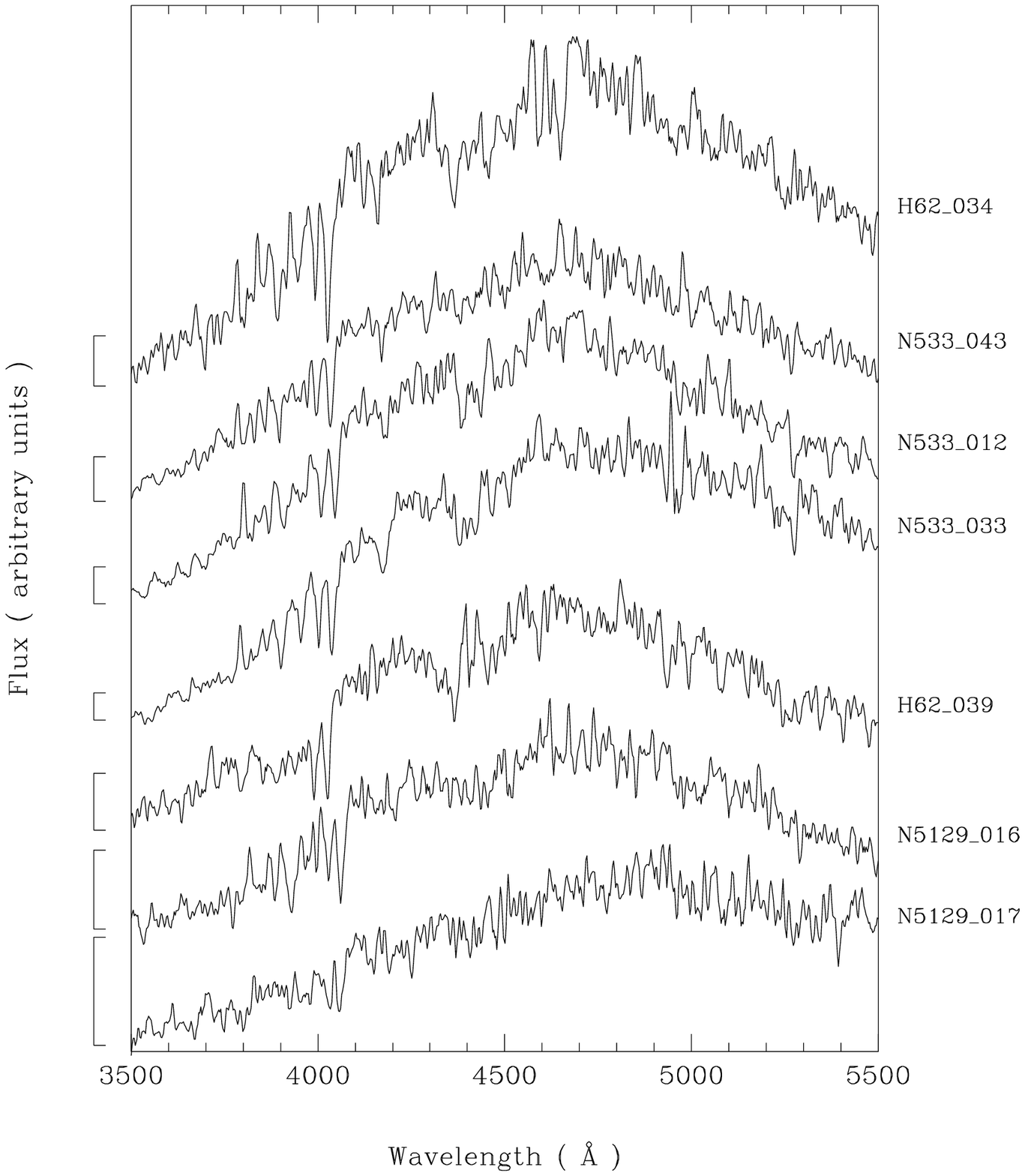}
\caption{{\it cont. }}
\end{figure}
\clearpage
\begin{figure}
\plotone{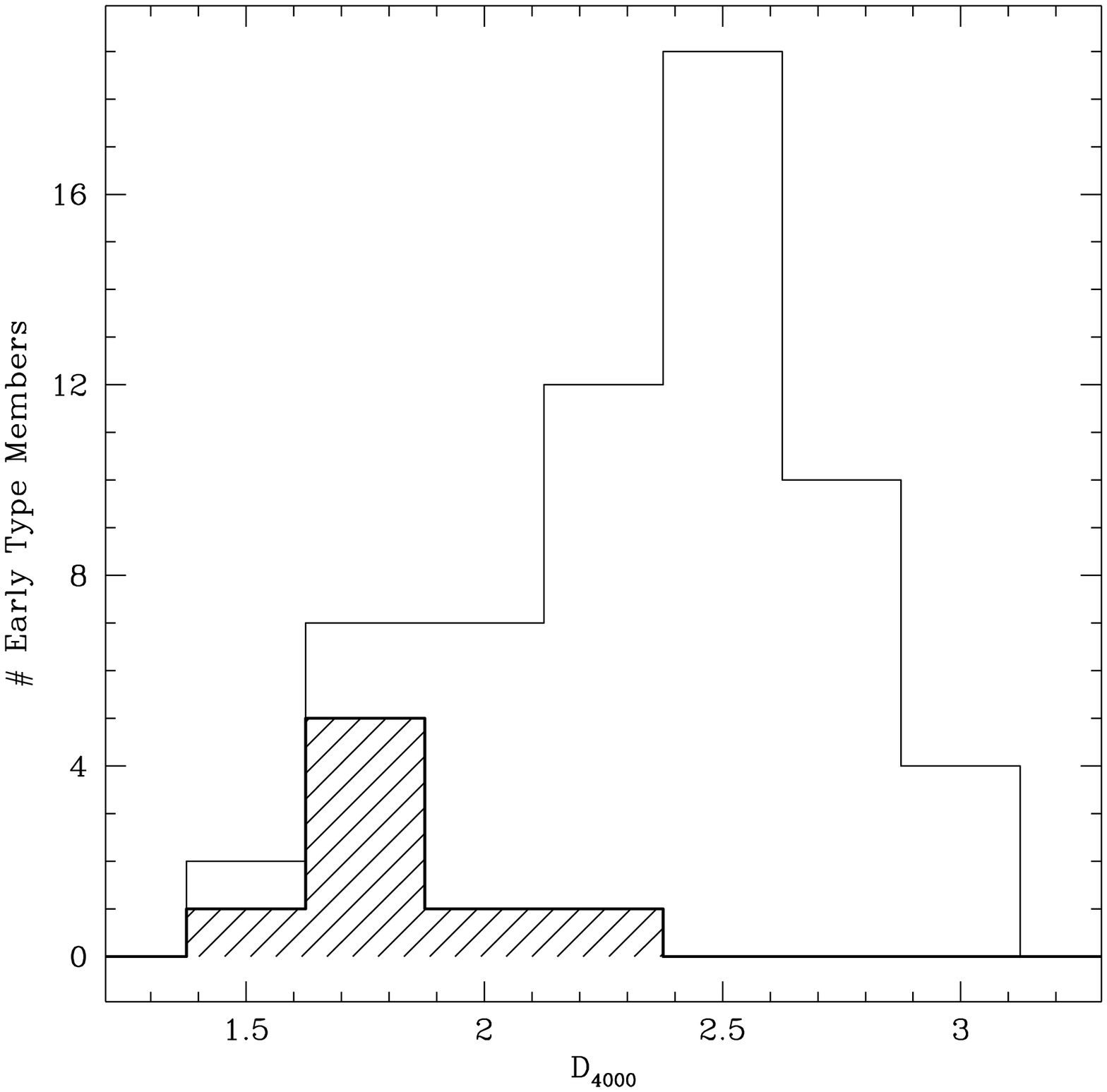}
\caption{}
\end{figure}
\clearpage
\end{document}